\documentclass[manuscript]{acmart}
\AtBeginDocument{%
  }

\usepackage{subcaption}
\usepackage{framed}
\usepackage[table]{xcolor}

\input{fig/tokens.tex}

\setcopyright{acmlicensed}
\copyrightyear{2018}
\acmYear{2018}
\acmDOI{XXXXXXX.XXXXXXX}
\acmConference[Conference acronym 'XX]{Make sure to enter the correct
  conference title from your rights confirmation email}{June 03--05,
  2018}{Woodstock, NY}
\acmISBN{978-1-4503-XXXX-X/2018/06}

\begin{document}

\title{Deflecting the Value Compass: Interacting with Large Language Models Temporarily Shifts Human Value Priorities Toward Personal Focus}

\author{Hasibur Rahman}
\affiliation{%
  \institution{Northeastern University}
  \city{Boston}
  \state{Massachusetts}
  \country{USA}}
  \email{rahman.has@northeastern.edu}

\author{Malak Sadek}
\affiliation{%
  \institution{Centre for Human-Inspired Artificial Intelligence (CHIA)}
  \institution{University of Cambridge}
  \city{Cambridge}
  \country{United Kingdom}}
  \email{mfzas2@cam.ac.uk}

\author{Smit Desai}
\authornote{Corresponding author}
\affiliation{
  \institution{Northeastern University}
  \city{Boston}
  \state{Massachusetts}
  \country{USA}}
\email{sm.desai@northeastern.edu}

\renewcommand{\shorttitle}{Deflecting the Value Compass}
\renewcommand{\shortauthors}{Rahman et al.}

\begin{abstract}
Large language models increasingly support decisions where values are in tension, yet little is known about whether interacting with them changes which values users prioritize. In a preregistered study, 200 U.S. adults interacted with ChatGPT, Claude, or Gemini as a thinking partner or read fixed AI-generated considerations. The prompt asked LLMs to support reasoning without recommending a decision and named no values. Participants advised people facing real dilemmas and completed parallel PVQ-RR forms before, immediately after, and one task later. Each LLM condition temporarily shifted value priorities toward personal focus relative to the control ($d=0.37$--$0.51$), primarily through increased Self-Enhancement. Participants' advice retained words and meaning from their exchanges. Thus, a brief LLM interaction that neither targets values nor seeks to persuade can reorient values active during judgment without detectable convergence in value directions or advice.
\end{abstract}



\begin{teaserfigure}
 \includegraphics[width=\textwidth]{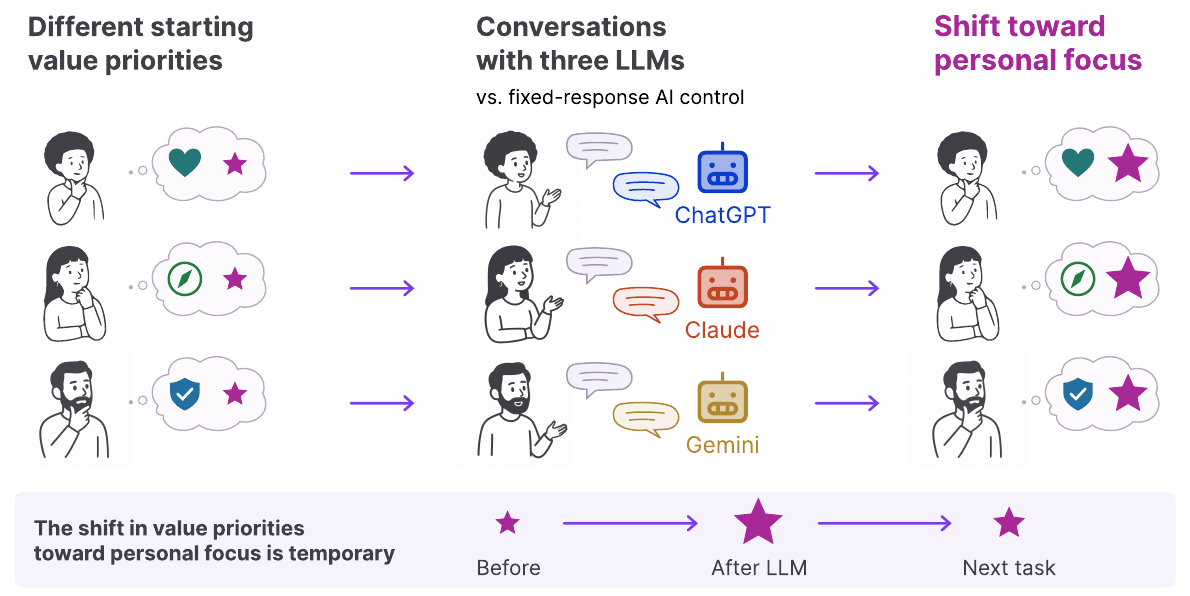}
  \caption{Overview of our three-phase study ($N=200$). Participants wrote advice and reported value priorities for a different dilemma in each phase. Participants were randomly assigned to converse with ChatGPT, Claude, or Gemini, or read fixed AI-generated text in Phase-2. Phase-1 and Phase-3 involved no AI. Conversations shifted priorities toward personal relative to social focus (larger stars) in Phase-2; the shift receded by Phase-3.}
  \Description{Three columns connected by purple arrows show different starting value priorities, conversations with three large language models, and a shift toward personal focus. Three rows depict people with distinct appearances. In the left column, each person thinks about a heart, compass, or shield alongside a small plum star. In the middle column, the same people gesture toward blue, orange, and gold robots labeled ChatGPT, Claude, and Gemini. Speech bubbles represent text-based exchanges; each robot's reply bubble matches its color. A black subtitle reads: vs. fixed-response AI control. This control involved reading fixed AI-generated considerations without follow-up interaction and is not depicted as a fourth row. In the right column, the people resume thinking, with their original symbols alongside larger stars. The stars indicate greater personal relative to social focus, not a particular value category. A pale-purple strip below states that the shift in value priorities toward personal focus is temporary. Small, large, and smaller stars labeled Before, After LLM, and Next task show the shift and its recession. The illustration summarizes a group-level pattern; it is not a quantitative plot or a mapping of individual participants to model conditions.}
  \label{fig:teaser}
\end{teaserfigure}

\maketitle

\section{Introduction}
\label{sec:intro}

People increasingly turn to large language models to reason through decisions in which their values pull in different directions. Someone is offered a better-paid job in another city, while their partner prefers the city where they currently live. The decision brings personal advancement into tension with commitments to a close other. They ask an LLM what is worth weighing. The model presents considerations on both sides and leaves the choice to them. Yet apparent balance can conceal unequal emphasis: the exchange may dwell on career opportunity or devote more attention to the relationship. Conversations like this now happen at enormous scale \cite{chatterji2025}. If a short exchange changes the relative importance of those considerations, it has shifted the values guiding the decision without ever making a recommendation.

Values are trans-situational goals whose importance is understood relative to other values, rather than at fixed levels \cite{schwartz2012overview}. A person may care about both personal advancement and their relationship; their relative priority helps guide the choice \cite{sagiv2017values}. Schwartz's refined theory arranges nineteen values on a circular motivational continuum and aggregates them into four higher-order values: Openness to Change, Self-Enhancement, Conservation, and Self-Transcendence. These four can be further grouped along a personal-social division. Personal focus combines Openness to Change and Self-Enhancement, capturing values that regulate the expression of personal interests. Social focus combines Conservation and Self-Transcendence, capturing values that regulate one's relations with others \cite{schwartz2012refined}. The decision above can therefore be understood partly along this axis: how much weight to give personal advancement relative to commitments to a close other.

A person's broader value hierarchy generally changes slowly over years \cite{bardi2011dual}. The values active during a particular judgment, however, are responsive to context and especially predictive of the choice made at that moment \cite{verplanken2002motivated,sagiv2017values}. Brief interventions can move reported value priorities \cite{maio2009changing,russo2022values}, as can the act of asking people to reflect on them \cite{bernard2003introspection}. A short interaction could therefore reorient the values active during deliberation without producing an enduring change in the person's broader hierarchy.

There is also reason to examine the direction of any such movement. Mainstream LLMs have been found to express a narrow set of human values aligned with Western, Educated, Industrialized, Rich and Democratic (WEIRD) populations \cite{tao2024}. More broadly, research on AI development has shown how technical systems can reproduce the priorities of those who build them \cite{birhane2022,holstein2019}. These findings characterize patterns in system outputs. Whether ordinary interaction with these systems reorients the value priorities people bring to a decision remains open.

Additionally, prior work has established that writing alongside an LLM configured to favor one side of a debate produces text that leans the model's way and, afterward, reported attitudes that lean the same way \cite{jakesch2023cowriting}. Advice from a model on moral dilemmas can similarly shift subsequent judgments toward the position it expresses \cite{krugel2023moral}. In both cases, the model advances a discernible direction. A thinking partner occupies a subtler role. Such systems are intended to extend users' reasoning through feedback and reflective questions while leaving the eventual judgment to them \cite{reicherts2025extendai,khadar2025socratic,tarvirdians2026reflectimate}. Even within that role, an LLM selects which considerations to raise, how to frame them, and how much attention each receives.

This distinction is important because prior research has primarily examined changes in what people think, say, or produce about a particular issue. Attitudes and outputs are object-specific, whereas values are abstract and help organize how people resolve conflicts across situations \cite{schwartz1992universals}. A temporary reorientation of the values active during judgment could therefore shape a decision even when the LLM neither advocates a position nor settles the choice. We investigate whether this occurs, how long it lasts, and whether a common directional movement makes participants' values or subsequent advice more alike by asking the following questions:

\begin{itemize}
    \item \textbf{RQ1.} Does interacting with an LLM shift participants' reported value priorities relative to a non-interactive AI control, and does any shift persist into a subsequent task without the LLM?

    \item \textbf{RQ2.} How does interacting with an LLM affect the direction, magnitude, and dispersion of participants' value profiles?

    \item \textbf{RQ3.} How do participants perceive the LLMs, do these perceptions differ across providers, and do they predict variation in value-priority shifts?

    \item \textbf{RQ4.} What linguistic and semantic traces of participants' LLM exchanges remain in their subsequent advice, and does that advice converge across participants?
\end{itemize}

We report a preregistered experiment with 200 U.S. adults (Figure~\ref{fig:teaser}). Across three phases, participants read different real dilemmas contributed by advice seekers, wrote the advice they would give, and completed parallel forms of Schwartz's PVQ-RR. The experimental conditions diverged during Phase-2. Participants assigned to  \arm{ChatGPT}, \arm{Claude}, or  \arm{Gemini} interacted with the model for up to ten minutes under the same thinking-partner prompt. The prompt named no values and prohibited the LLM from advocating for values, recommending an action, or settling the dilemma. Participants in the control read fixed AI-generated considerations synthesized from the same three models but could not send follow-up messages. All participants then wrote advice and completed the same measures. Phase-3 introduced another dilemma without an LLM in any condition, allowing us to examine what remained by the following task. The advice-giving format held the value conflict constant while psychological distance foregrounded abstract, desirability-based considerations \cite{trope2010,danziger2012idealistic}.

Immediately after Phase-2, participants in each LLM condition shifted toward personal focus by $0.31$--$0.43$ scale points relative to the control (Glass $d=0.37$--$0.51$), primarily through increased Self-Enhancement. The four higher-order values rotated together, and the three LLM conditions moved along closely aligned bearings. By Phase-3, the contrasts with the control had shrunk to near zero and none survived correction. This common movement occurred without detectable convergence in value directions or advice. Their advice nevertheless retained words and meaning specific to their own exchanges. Participants' perceptions of the LLMs did not differ across providers and did not predict the shift.

These results make three contributions:

\begin{itemize}
    \item \textbf{Empirically,} we provide controlled evidence that brief interaction with three commercial LLMs under a non-directive thinking-partner role can temporarily shift reported value priorities toward personal focus beyond exposure to fixed AI-generated considerations.

    \item \textbf{Conceptually,} we identify a form of LLM influence in which participants move in a common direction and show exchange-specific uptake without detectable convergence in value directions or advice.

    \item \textbf{For the evaluation of human--AI interaction,} we show that timing and outcome measures materially affect whether influence is visible. Immediate value measures captured a shift that had largely receded by the following task, while participants' perceptions of the systems did not track who moved.
\end{itemize}

\section{Related Work}

\subsection{Defining and Measuring Human Values}
\label{sec:rw-values}

A significant body of work has explored the interplay between various technologies and human values \cite{umbrello2019, burken2023, vernim2022, sadek2025vsca}. The nature of human values is heavily debated, with numerous definitions \cite{taylor1977, vermaas2020, kroes2020} and operationalizations \cite{poel2020, klingefjord_what_2024} available. We take values to mean concepts that people consider important in their lives, govern how they wish and choose to live, and help them distinguish between positive and negative outcomes and behaviors \cite{friedmanbook}.
 
While values themselves are enduring, value expressions and priorities are flexible \cite{bardi2014, maio2009changing}. Prioritized values differ across countries and cultures \cite{schwartz2004}; even within narrower groups, technology-related values differ dramatically by factors such as job roles \cite{jakesch2022}, technical expertise \cite{birhane2022, holstein2019}, and demographic factors \cite{sadek2026cui}.  More microscopically, at the individual level, values can change across situations \cite{gornemann2022}, throughout their lives \cite{nathan2008}, and in response to certain triggers or interventions \cite{desmet2015}, including simply being asked about values \cite{sadek2025vsca}.

We use \textit{value priorities} to refer to the expression of values activated during a particular judgment. A person's underlying value hierarchy is usually stable and only changes gradually over the years \cite{bardi2009structure, bardi2011dual}. Specific value priorities in a given moment, however, are more context-sensitive. Values influence a choice when they are activated, and the values active at that moment best predict the choice \cite{verplanken2002motivated, sagiv2017values}. Social cognition explains why such expression should shift after a brief exposure and then fade. A construct made accessible by recent exposure shapes how later ambiguous information is interpreted, and its influence weakens as the interval since exposure lengthens \cite{higgins1977category, srull1979accessibility, higgins1985priming, hong2000multicultural}. Prior studies have therefore measured reported priorities immediately after a value-relevant task to capture short-term changes in their ordering \cite{maio2009changing, russo2022values}.

A review of 25 value-manipulation experiments confirms that reported value priorities can be moved with small-to-medium effects \cite{russo2022values}. After one 30-minute intervention, effects were still detectable four weeks later \cite{arieli2014benevolence}. Additionally, those manipulations were directive in that they asked participants to analyze reasons for named values, primed particular values, or induced dissatisfaction through self-confrontation or mortality salience \cite{maio1998truisms, bernard2003introspection, maio2009changing, ye2019culturalpriming, russo2022values}. These findings establish that targeted interventions can shift reported priorities, but leave open whether a conversation with an LLM that neither names values nor advocates for them can do so. This gap motivates RQ1 to determine whether such a task shifts priorities at all. 

The Portrait Values Questionnaire (PVQ) is among the most established instruments for measuring values and detecting value-priority shifts. Its revised 57-item form, the PVQ-RR \cite{schwartz2022pvqrr}, operationalizes Schwartz's refined theory of basic individual values \cite{schwartz2012refined}, partitioning the ten basic values of the original theory \cite{schwartz1992universals} into nineteen narrower values arranged on the same circular scale. Values adjacent on the circle express compatible motivations and tend to be endorsed together, while values opposite on the circle express conflicting motivations and trade off. The nineteen values aggregate into four higher-order values forming two dimensions: openness to change against conservation, and self-enhancement against self-transcendence \cite{schwartz1992universals}. A personal--social division crosses the circle, separating values that regulate one's own interests from those that regulate relations with others \cite{schwartz2012refined}. This structure lets us ask not only whether priorities shift, but also which motivations gain relative importance. For RQ1, the personal--social division lets us examine the direction of any shift.


\subsubsection{Whose Values? Value Skew in Large Language Models}
\label{sec:rw-skew}

Values are not universal \cite{poel2020, varanasi2023}, and imposing one set of values on diverse contexts and users has been widely criticized \cite{wallach2009, vera2019, palmer2023, sorensen2024pluralistic}. Nevertheless, AI systems have been found to embody a narrow set of values \cite{jakesch2022, sadek2024responsible}, particularly those of their creators \cite{birhane2022, holstein2019}. Current alignment practices further narrow them, with standard procedures found to reduce distributional pluralism \cite{sorensen2024pluralistic}, and people supplying preference data disagree widely across cultures \cite{kirk2024prism}. 
 
For LLMs, the skew is consistently in one direction. Audits across value frameworks, languages, personas, and cross-cultural dilemmas find that commercial models align with Western, Educated, Industrialized, Rich, and Democratic (WEIRD) populations \cite{birhane2022, kazemi2024, tao2024, cao2023, zhou2025, sambasivan2021, wang2024cdeval, karinshak2024llmglobe, alkhamissi2024cultural, lu2025cultural, rahman2026ccd}, and models that complete the PVQ-RR themselves diverge from human baselines \cite{hadar2024}. These are descriptions of outputs, which vary with wording and steering \cite{rottger2024politicalcompass, santurkar2023opinions}, but the WEIRD-skew persists across paraphrases, translations, and personas \cite{moore2024valueconsistency, lee2026valueinertia}, predicts model behavior \cite{yao2024valuefulcra}, and surfaces in deployment, where the values a model expresses are task-dependent and partly mirror the user's own \cite{huang2025valueswild}.
 
Output skew alone does not establish whether users adopt similar priorities from a single conversation when LLMs neither advocate nor name values. LLMs can influence perceptions \cite{li2026personality} and opinions \cite{jakesch2023cowriting}, while value priorities respond to situations and experiences \cite{maio2009changing, desmet2015, gornemann2022}. Together, these findings motivate RQ1 to ask whether priorities shift after a non-directive conversation and whether they shift toward personal rather than social focus. 


\subsection{The Impacts of Large Language Models on Users}
\label{sec:rw-shifts}

General concerns about the impacts of AI systems \cite{janowicz2025} are amplified by LLMs' social and interactive nature \cite{shahid2026}. LLMs can persuade users \cite{salvi2024, hackenburg2025levers, gallegos2026}, deceive users \cite{hagendorff2024, Yeo_Jin_Noh_Shin_Kang_Heo_Chung_Hyun_Han_2026}, and foster unhealthy reliance \cite{fang2025, collins2024} in users. Participants who wrote with an LLM that was biased toward one side of a debate and later reported attitudes leaning that way \cite{jakesch2023cowriting}. Most participants were unaware of the influence, warnings did not mitigate it, and an interactive assistant moved attitudes more than the same suggestions shown as static text \cite{Williams-Ceci_Jakesch_Bhat_Kadoma_Zalmanson_Naaman_2026}. These findings leave open whether influence found with a deliberately biased assistant extends to a non-directive thinking partner.

Related to this work, \citet{teng2026moral} found that five-minute chats with a chatbot prompted to shift moral judgments shifted participants' judgments for two weeks, while neutral chats produced no change. Participants rated the chatbots similarly in likability and convincingness. This comparison establishes an effect of directed moral persuasion relative to an unrelated conversation; it leaves open what happens when an LLM supports deliberation about a dilemma without an assigned persuasive direction. It also leaves a distinction between verdict changes and changes in relative value priorities that may inform judgments across situations. RQ1 examines the latter after a thinking-partner conversation.

Research shows that these impacts outlast an LLM interaction. For example, LLM-generated messages durably persuaded users politically at scale \cite{hackenburg2025levers} and extended dialogues with LLMs durably decreased users' belief in conspiracy theories \cite{costello2024conspiracy}. These impacts also carry over into tasks without the system. For example, self-confidence converged with an AI advisor's expressed confidence and stayed aligned in later unassisted decisions \cite{li2025confidence}, self-reported personality moved toward a chatbot's traits \cite{li2026personality}, chatbot interaction changed what users then disclosed to a human professional \cite{leemediator2020, leedisclosure2020}, and AI influence spilled into unrelated human-human interactions, described as a ``social forcefield'' beyond the exchange itself \cite{riedl2026, harrell2025}. Such carryover predates LLMs, as when people's behavior shifts to match a digital avatar representing them \cite{proteus2007}. This evidence motivates RQ1's follow-up question of whether a value-priority shift persists during a subsequent advice task without the LLM.

The following sections situate the LLM's role in research on conversational reflection and deliberation, then examine how an exchange can leave traces in what users write. In our study, the LLM supported the participant's reasoning, and the participant authored advice for the person facing the dilemma.

\subsubsection{LLM Support for Reflection and Deliberation}
\label{sec:rw-advice}

Research on LLM thinking partners examines how systems can complement people's reasoning through interaction \cite{collins2024thoughtpartners}. Within HCI, LLMs have been used to help users articulate assumptions and reconsider their reasoning. ProberBot prompted reflection during investment decisions through questions \cite{reicherts2022proberbot}, while ExtendAI embedded feedback in users' rationales and compared this support with recommendation-based assistance \cite{reicherts2025extendai}. More recently, a Socratic LLM questioned annotators' reasoning while leaving conclusions to them \cite{khadar2025socratic}, and Reflecti-Mate used adaptive questions to support reflection on personal decisions without directive advice \cite{tarvirdians2026reflectimate}. These systems situate LLM assistance within the process through which users develop a judgment.

This setting raises a question about what users prioritize as they reason. Communication can affect the relative importance assigned to considerations in a judgment \cite{chong2007}, suggesting that deliberative support could shift priorities even without a supplied recommendation. RQ1 examines whether reported value priorities change following a thinking-partner exchange with no assigned persuasive direction, relative to a fixed presentation of AI-generated considerations.


Advice-giving offers a setting for such deliberation. Everyday dilemmas drawn from online advice communities expose conflicts between values \cite{lourie2021,forbes2020}. Advising another person creates psychological distance that foregrounds abstract, desirability-based considerations \cite{trope2010,danziger2012idealistic}. Written advice makes the considerations behind a judgment available for analysis. When developed through conversation with an LLM, this reasoning may also retain language and ideas from the exchange, connecting deliberative support to research on LLMs' influence on users' writing. Advice-giving tasks, therefore, let us examine whether LLM-supported deliberation shifts reported value priorities (RQ1) and leaves detectable traces in the reasoning participants subsequently express (RQ4).

\subsubsection{The impacts of LLMs on users' writing outputs}
\label{sec:rw-writing}

LLMs' impacts also travel through language. Speakers converge linguistically with their partners automatically during dialogue \cite{brennan1996, pickering2004dialogue}, including with computers \cite{branigan2010}. Writing with LLMs shifts opinions \cite{jakesch2023cowriting}, self-concept \cite{li2026personality}, and style \cite{agarwal2025}. Because LLMs default to Western conventions, this convergence pulls non-Western users' writing toward those norms unless they prompt against it \cite{agarwal2025}, a hidden cost for the Majority World that runs from stereotype reinforcement and global power imbalances to less usable and more harmful systems \cite{oyemike2025, safir025, naous2024, durmus2024, shelby2023, AIcolonialism, okolo2024, zhou2025}. At the collective level it flattens output, where stories written with AI ideas resemble one another more than stories written without \cite{doshi2024generative}, creative writing and crowd work homogenize \cite{agarwal2025, veselovsky2025}, and outputs converge toward an algorithmic monoculture \cite{kleinberg2021, irani2010}.
 
This convergence may also feed back on users through their own language. Communicators who tailor a message afterward remember and evaluate its subject in line with what they wrote \cite{higgins1978saying}, and people infer their attitudes partly from their own behavior \cite{bem1972selfperception}. When writing precedes measurement, model wording reproduced in their own text is a candidate channel for a shift, distinct from reading the model's output. RQ4 asks whether this channel is present, whether it carries meaning as well as words, and whether adoption is uniform across users or specific to each conversation. The homogenization findings also motivate RQ2, which distinguishes convergence in profile direction from changes in the spread of profile magnitudes, complementing RQ1.

\subsubsection{User perceptions as mediators of LLM impacts}
\label{sec:rw-partner}

Users apply human social rules to computers while knowing they are not people \cite{nass2000, reeves1996}. The metaphor introducing an LLM as a collaborator versus a tool changes how people evaluate and use it, even when its behavior is held constant \cite{desai2023, desai2024, desai2026, khadpe2020}, and an LLM's linguistic persona shapes trust, likeability, and adoption \cite{Genc_Gu_Degachi_Niforatos_Chandrasegaran_Verma_2026, rahman2026vibecheck}. Whether an LLM acts as advisor, peer, or decision-maker changes how far users' self-confidence follows it \cite{li2025confidence}, and systems that extend users' reasoning without a verdict behave differently from those that recommend one \cite{reicherts2025extendai, danry2023askme}.

Despite this, whether favorable user perceptions make a model more impactful is still debated. Perception measures did not distinguish a persuasive LLM from a neutral one \cite{teng2026moral}, and persuasion survived disclosure \cite{gallegos2026}. These mixed findings motivate RQ3, asking how participants perceive their conversational LLM partner, whether perceptions differ across providers, and whether more favorable ratings predict larger value-priority shifts.

\section{Study Method}
We ran a preregistered Institutional Review Board-approved study with 200 adults to test whether brief LLM conversation changes personal value priorities. Participants were randomly assigned to \arm{ChatGPT}, \arm{Claude}, \arm{Gemini}, or \arm{Baseline}. In each of three phases, they read an advice-seeking dilemma, wrote advice, and answered a values questionnaire. Conditions differed only in Phase-2: conversational participants chatted with a live LLM introduced as a thinking partner before writing advice; \arm{Baseline} read a fixed AI-generated response to the same dilemma, synthesized from the same three LLMs' outputs and delivered by the same assistant in the same interface (\S\ref{sec:control}). \arm{Baseline} is therefore an AI condition without conversation; the Phase-2 contrast isolates the LLM interaction.

We first describe value measurement (\S\ref{sec:pvq}), then the dilemmas, advice task, and counterbalancing (\S\ref{sec:task}), the LLM conversation and static control (\S\ref{sec:conversation}), advice, perception, and intake measures (\S\ref{sec:measures}), and the procedure (\S\ref{sec:procedure}).

The LLM supported deliberation while participants decided what advice to give. We tested for shifts in reported value priorities and traces of the exchange in subsequent advice. Centered PVQ-RR scores were the primary self-report outcome; advice provided a complementary record of expressed reasoning. We compared conversational participants' Phase-2 advice with their own LLM's turns and a yoked conversation from the same model and dilemma. For \arm{Baseline}, the source was the fixed response read (\S\ref{sec:advice-measures}).


\subsection{Measuring Value Priorities}
\label{sec:pvq}
We operationalized value priority at the higher-order level, validated three parallel PVQ-RR forms, centered participants' ratings, and defined the change estimated by each condition contrast.

\subsubsection{Construct and primary axis}
\label{sec:construct-level}
 

The PVQ-RR represents 19 refined values arranged in a circular motivational continuum. Schwartz groups these values into four higher-order dimensions: Openness to Change, Self-Enhancement, Conservation, and Self-Transcendence \cite{schwartz2012overview,schwartz2012refined}. We use these four dimensions throughout because they define the value profile and the personal-versus-social contrast examined in our RQs. We did not analyze the 19 refined values separately. The rotation decomposition likewise uses the four higher-order dimensions (Section~\ref{sec:analysis}).

\begin{figure*}[ht]
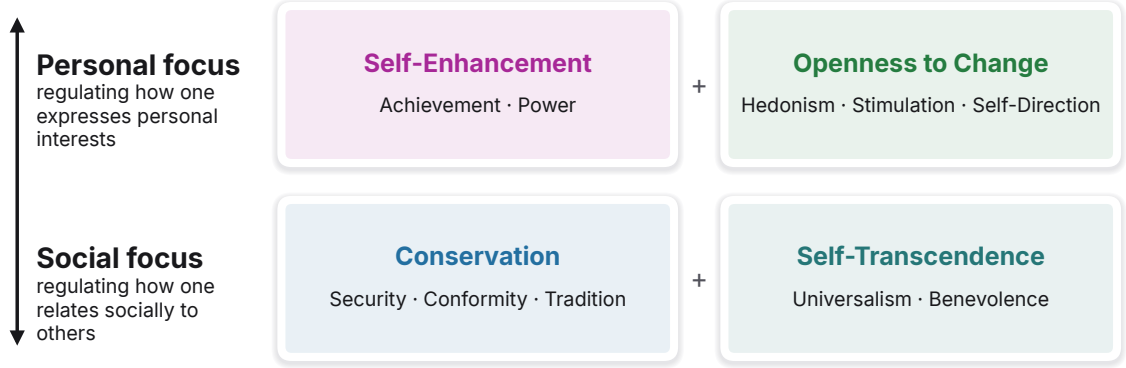

\centering
\figincl{value_space}
\caption{\textbf{The personal-versus-social value axis used as this study's primary outcome.} Personal focus combines Openness to Change and Self-Enhancement; social focus combines Conservation and Self-Transcendence. Each dilemma supports a defensible recommendation aligned with either pole. The cards list the ten basic values nested within the four higher-order groups. Adapted from Schwartz's value circle and its personal-focus and social-focus grouping \cite  {schwartz2012overview,schwartz2012refined}.}
\Description{Two rows of value cards with a vertical double-headed arrow running down the left edge. The label Personal focus heads the upper row. The row holds two cards: Self-Enhancement (Achievement and Power) and Openness to Change (Hedonism, Stimulation, and Self-Direction). The label Social focus heads the lower row. Its gloss is regulating how a person relates socially to others. The row holds two cards: Conservation (Security, Conformity, and Tradition) and Self-Transcendence (Universalism and Benevolence). A plus sign sits between the two cards in each row. The arrow spans both rows, marking them as the two ends of one dimension.}
\label{fig:value}
\end{figure*}

These dimensions define the built-in contrast in the dilemmas. Openness to Change and Self-Enhancement form the personal pole, whereas Conservation and Self-Transcendence form the social pole \cite{schwartz2012overview,schwartz2022pvqrr}. Their difference is our primary outcome because it connects the value measure to the personal-versus-social choice presented by each dilemma (Figure \ref{fig:value}). LLM-side evidence of value skewness motivates personal-versus-social comparison (\S\ref{sec:rw-skew}).



\subsubsection{PVQ-RR and its parallel forms}
To measure the four higher-order dimensions across three phases without repeating portraits, we constructed parallel forms of the PVQ-RR. The full instrument has 57 portraits, three per refined value~\cite{schwartz2012refined,schwartz2022pvqrr}, rated from 1 (not like me at all) to 6 (very much like me).

We divided the instrument into blocks A (items 1--19), B (20--38), and C (39--57), each with one item per refined value. Each block measured Openness to Change with four items (including Hedonism), Self-Enhancement with three, and Conservation and Self-Transcendence with five each. Face and Humility entered the centering mean, but not these scores, because each spans neighboring dimensions. Participants completed one block per phase, with no repeated portraits and item order randomized within each block and phase. Published reliability estimates use three items per refined value and do not validate these one-item-per-value blocks~\cite{schwartz2022pvqrr}; we therefore pre-validated them at the analyzed higher-order level (\S\ref{sec:construct-level}). All outcomes are multi-item composites: higher-order scores average three to five items, and the personal-focus axis compares seven with ten. A sensitivity check assigning Face to Conservation and Humility to Self-Transcendence yielded a nearly identical axis ($r=.985$), larger Phase-2 contrasts, and Phase-3 contrasts still near zero.

\textbf{\textit{The pre-study }}recruited 100 people under the study criteria. They completed all 57 items in a randomized block order, one block at a time. Block higher-order scores correlated $0.88$--$0.97$ with full-instrument scores. All 12 block-to-full and all 12 block-to-block comparisons were equivalent within $0.5$ scale points by two one-sided tests~\cite{lakens2017equivalence}. Single-block intraclass correlations were $0.66$--$0.84$ (mean $0.76$), and three-block correlations were $0.85$--$0.94$~\cite{shrout1979icc}. Mean pairwise correlation was $0.79$; full-instrument Cronbach's $\alpha$ was $0.86$--$0.93$. These results support the 19-item parallel forms at the higher-order level (Table~\ref{tab:parallel-forms}).

Blocks retained small mean offsets, reaching $0.37$ scale points for Openness to Change and Self-Enhancement; Self-Transcendence showed no block effect ($p=.589$). We counterbalanced block order independently of condition and included block as a covariate in every model. Thus, offsets were not systematically tied to phase or condition and were adjusted in condition-by-phase contrasts (\S\ref{sec:estimand}).

\subsubsection{From ratings to priorities}

Parallel forms make phases comparable, but value priorities also require separating relative importance from response-scale use: rating every item 5 or every item 3 expresses the same priorities. Schwartz therefore recommends subtracting each person's mean rating (MRAT) across all value items to express relative importance rather than general scale use \cite{schwartz2012overview}. For each phase's parallel form, we subtracted its MRAT across all 19 responses, including Face and Humility, from its four higher-order scores~\cite{schwartz2012overview}. We also report uncentered group means and MRAT shifts because centered scores can decrease as MRAT increases, even when raw group means change little.

Centering makes the four group scores compositional. They sum to approximately zero, so a rise in one mechanically pushes others down \cite{aitchison1982compositional}. This dependence suits relative priorities, but reduced between-person spread in centered profiles cannot establish declining absolute endorsement. We therefore separate profile direction from magnitude for RQ2 and repeat the spread analysis with uncentered scores and MRAT (Appendix~\ref{app:supporting}).

We computed the primary personal-focus axis from the centered scores. Personal focus is the item-count-weighted mean of Openness to Change (OC) and Self-Enhancement (SE); social focus is the item-count-weighted mean of Conservation (CO) and Self-Transcendence (ST). The axis is their difference, positive toward personal focus:
\[
\frac{4\,\mathrm{OC} + 3\,\mathrm{SE}}{7}
-
\frac{5\,\mathrm{CO} + 5\,\mathrm{ST}}{10}.
\]
The weights keep both composites on the raw PVQ-RR scale, so zero indicates equal weight on the two poles. Face and Humility enter only through MRAT.

\subsubsection{What the design estimates}
\label{sec:estimand}
Our estimand is a change rather than a score level. Let $\mu_{c,p}$ be the adjusted mean for condition $c$ at phase $p$, and $B$ be \arm{Baseline}. For each conversational condition and $p\in\{2,3\}$, \[ \tau_{c,p} = (\mu_{c,p}-\mu_{c,1}) - (\mu_{B,p}-\mu_{B,1}). \] Thus, $\tau_{c,p}$ is the Phase-1 change relative to the non-interactive AI control.

The contrast accounts for the shared dilemma--advice--PVQ-RR sequence, with counterbalancing and covariate adjustment addressing form and order differences (\S\ref{sec:counterbalancing}). We report $\tau_{c,p}$ as the primary result and each condition's change for context.

\subsection{The Dilemmas}
\label{sec:task}
Value priorities become most visible when judgments pit conflicting values against each other, and advice-giving foregrounds abstract, desirability-based considerations (\S\ref{sec:rw-advice}). Each dilemma, therefore, pitted personal against social focus, matching our primary axis. In every phase, participants read a dilemma in which someone else sought advice and wrote their advice.


\subsubsection{Designing and validating the dilemmas}
We used third-party advice tasks to hold the conflict constant while requiring a value-laden stance. Participants' own dilemmas would introduce personal stakes we could not equate across the sample; shared dilemmas made task conditions comparable for examining priorities and written reasoning. The inventory immediately followed the advice. \arm{Baseline} followed the same dilemma--advice--inventory path (\S\ref{sec:control}), so the conversational contrast estimates the increment over an AI-generated consideration.

We selected three dilemmas from VALACT-15K, a corpus of real advice-seeking Reddit posts mapped onto Schwartz's values~\cite{huang2026valact}. Selection required comparable reading demands, the same personal-versus-social conflict across different everyday decisions, and explicit mention of considerations aligned with both poles. We first retained posts of 200--600 words; an LLM shortlisted 30 in which the four higher-order values supported distinct actions. Three researcher consensus rounds narrowed these to 12, then 6, then 3, excluding posts involving sexual content, potentially distressing material, or physical or psychological abuse. The retained posts preserved the same value structure across career, family, and household decisions.

The retained dilemmas preserved each poster's first-person voice. We removed identifying details, including ethnicity markers.

\textbf{Job} asks whether a person should accept a better-paid, more senior position in another city or remain in a preferred city with a partner who likes the current city but dislikes his job.

\textbf{Family} concerns an 18-year-old with a sports scholarship who wants to become a personal trainer. His stepfather wants him to
study law to preserve the family's reputation, while his mother works night shifts and wants peace at home.

\textbf{Move-out} concerns a 23-year-old who pays half the household rent and provides most of the care for a younger sibling with a learning disability. He has signed a tenancy agreement two hours away while the family's finances deteriorate.

Together, the dilemmas vary in decision context while holding the personal-versus-social conflict constant. We did not score the recommendation direction; instead, the advice task required active engagement with each dilemma and provided RQ4's behavioral record.

\subsubsection{The advice task}
\label{sec:counterbalancing}

At each phase, participants wrote advice in a plain editor with pasting and drag-and-drop disabled. Responses required at least 30 words, with no upper limit. This minimum followed evidence that Divergent Semantic Integration (DSI) scores stabilize for texts of roughly 30--50 words \cite{johnson2023dsi}. 

We counterbalanced dilemma and PVQ-RR block order with six Williams-balanced Graeco-Latin square sequences \cite{williams1949designs}. Each participant encountered every dilemma and block once; each appeared in every phase across sequences, with reversed sequences balancing first-order carryover. Item order was randomized within each block and phase. We rotated sequences within each between-participant condition. Because 50 participants cannot divide equally among six sequences, models adjusted for block, dilemma, and sequence (\S\ref{sec:analysis}).

\subsection{The LLM Conversation and the Non-Interactive AI Control}
\label{sec:conversation}

We designed the interaction to support participants' deliberation. An LLM prompted to stay neutral can still move preferences after a short exchange \cite{wise2025chatbot}, so the thinking-partner stance controls the manipulation. One prompt, identical across \arm{ChatGPT}, \arm{Claude}, and \arm{Gemini}, instructed models to support reasoning while leaving recommendations to participants, with no target recommendation to promote.

Within these role constraints, the conversation was unscripted after a shared opener and without a response template \cite{yun2025format}. Content and emphasis could therefore develop organically with the participant's contributions in each exchange, so that similarity across exchanges and subsequent advice could be examined empirically.

\subsubsection{LLM conversation}
The shared prompt prohibited advocating for values, recommending actions, or settling the dilemma. The system prompt varied only in the dilemma text; replies remained unscripted. The role clause reads:

\begin{quote}\small\itshape
You are a thinking partner helping the reader reason through what to recommend, never telling them what to recommend and never settling it for them.
\end{quote}

To avoid eliciting agreement with an initial participant lean \cite{sharma2024sycophancy,cheng2026sycophantic,jain2026sycophancy}, every conversation began with the same uneditable click-to-send message:

\begin{quote}\small\itshape
I just read this situation and I have to write up what I'd recommend. Before I do, what are the main things worth weighing here?
\end{quote}

Participants then typed freely with the dilemma visible. The interface identified the model only as an AI assistant, subtitled ``Thinking partner''. The continue button appeared after five participant messages, or when the ten-minute cap elapsed.

\begin{figure*}[htbp]
\centering
\figincl[width=.8\textwidth]{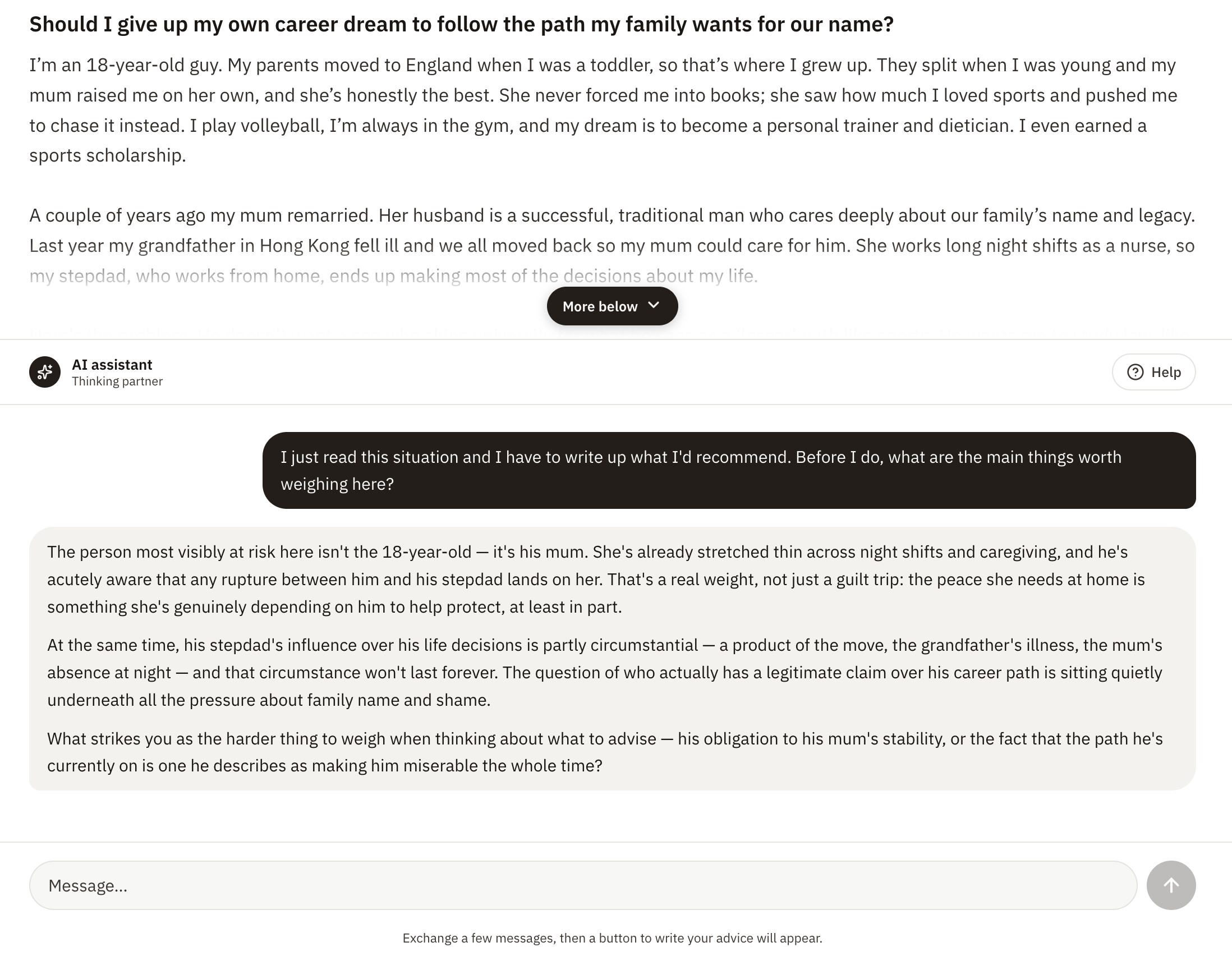}
\caption{\textbf{Study interface in a conversational condition.} The scrollable dilemma remains above the exchange. A shared ``AI assistant'' header and a ``Thinking partner'' subtitle conceal the provider's identity. \textit{For Baseline, the text input was disabled.}}
\Description{The upper panel shows the advice-seeking post and a More below scroll indicator. Beneath it, the AI assistant header includes the Thinking partner subtitle and a Help button. A dark participant bubble contains the shared opening question; a light assistant bubble contains considerations and a follow-up question. A message field and send button appear at the bottom, followed by a note that the advice button will appear after a few messages.}
\label{fig:ui}
\end{figure*}

Generation settings were identical across the three models. Calls went through OpenRouter to \texttt{openai/gpt-5.4}, \texttt{anthropic/claude-sonnet-4.6}, and \texttt{google/gemini-3-flash-preview}, with temperature $0.7$. The request set no token cap, \texttt{top\_p}, or seed. Median first replies ranged from 87 to 191 words across providers (\S\ref{res:manipulation}).

\subsubsection{The non-interactive AI control}
\label{sec:control}

\arm{Baseline} retained exposure to AI-written text, AI attribution, and the interface, but offered no exchange (\S\ref{sec:rw-writing} and \S\ref{sec:rw-advice}). For each dilemma, the same three models restated the facts and considerations already expressed in the dilemmas. This changed their presentation without adding new considerations or advocacy.

An LLM synthesized overlapping content into one fixed response per dilemma. Pairwise cosine similarities among the source responses were $0.836$ for \textbf{Job}, $0.724$ for \textbf{Family}, and $0.763$ for \textbf{Move-out}. Each synthesis spanned $345$--$359$ words and had cosine similarity above $0.85$ to every source. The control was thus synthesized across providers rather than provider-matched.

Participants read the response under the same AI-assistant header and ``Thinking partner'' subtitle, retaining the dilemma panel and message format but removing the composer. Everyone assigned a dilemma read the same response, then completed the same advice, PVQ-RR, and perception tasks as conversational participants. Thus, the contrast compared live, non-directive exchange with fixed presentation of existing considerations. Time on task outside the manipulation window did not differ across conditions (\S\ref{res:manipulation}). The fixed passage also served as a known source for validating the text-overlap measures (\S\ref{res:uptake}).

\subsection{Advice, Perception, and Intake Measures}
\label{sec:measures}

The PVQ-RR is the primary outcome measure. Advice texts address RQ4, perception measures address RQ3, and intake measures describe the sample. 

\subsubsection{Advice texts}
\label{sec:advice-measures}
We measured traces of the AI source, semantic focus, and similarity of the written advice across participants. Let $a_i$ be participant $i$'s Phase-2 advice and $s_i$ the concatenated messages they received. Bold symbols denote $\ell_2$-normalized embeddings; $\operatorname{unit}(\mathbf{v})$ normalizes a vector.

Each advice text was compared with its own source and a yoked source from the same provider and dilemma. Within each cell, participants were paired in data order with the next participant, wrapping around so each source served once as a yoke. For \arm{Baseline}, we compared the passage read with one for an unseen dilemma as a positive control; identical fixed passages would yield no contrast.

We measured resemblance lexically and semantically. Lexical overlap was
\[
m_{n}(a, s) \;=\; \frac{\lvert G_{n}(a) \cap G_{n}(s) \rvert}{\lvert G_{n}(a) \rvert},
\]
where $G_n(t)$ is the set of distinct content-word $n$-grams after lowercasing and removing a fixed English stopword list. We scored $n=1$, $2$, and $3$ and report the two-word score as phrase overlap. Semantic overlap was $m_{\cos}(a,s)=\mathbf{a}^{\top}\mathbf{s}$.

To distinguish uptake specific to an exchange from content shared across its provider and dilemma, we projected each source embedding onto a centroid of other sources and retained the orthogonal remainder:

\[
\mathbf{c}_i \;=\; \operatorname{unit}\!\Bigl(\frac{1}{k} \sum_{j \in K_i} \mathbf{s}_j\Bigr),
\qquad
\mathbf{u}_i \;=\; \operatorname{unit}\!\bigl(\mathbf{s}_i - (\mathbf{s}_i^{\top} \mathbf{c}_i)\, \mathbf{c}_i\bigr),
\]
Here, $K_i$ is a fixed random set of $k$ other participants in the same cell, $\mathbf{c}_i$ its centroid, and $\mathbf{u}_i$ the normalized residual. For comparable centroid precision, $k$ was one less than the smallest provider cell for each dilemma.

We measured semantic focus with Divergent Semantic Integration (DSI), adapted to sentence pairs \cite{johnson2023dsi}:
\[
\mathrm{DSI}(a) \;=\; \frac{2}{S(S-1)} \sum_{p < q} \bigl(1 - \mathbf{x}_{p}^{\top} \mathbf{x}_{q}\bigr),
\]
where $\mathbf{x}_{1}, \ldots, \mathbf{x}_{S}$ are the embeddings of the text's $S$ sentences. High DSI indicates more distinct ideas; low DSI indicates greater semantic focus. We split on sentence-final punctuation, used semicolons and line breaks when fewer than two segments remained, and retained segments
of at least three words.

Between-person similarity was each text's cosine distance from the other texts in the same condition, phase, and dilemma:
\[
d_i \;=\; 1 \;-\; \mathbf{a}_i^{\top}\, \operatorname{unit }\!\Bigl(\frac{1}{n_c-1} \sum_{j \neq i} \mathbf{a}_j\Bigr),
\]
where $n_c$ is the cell size. Lower $d_i$ indicates greater between-person similarity \cite{anderson2024homogenization}. The measure is undefined for cells with fewer than three texts.

All semantic measures used \texttt{text-embedding-3-small}. We also recorded advice length and sentence count. Vocabulary range used the moving-average type-token ratio over a 25-token window \cite{mccarthy2010mtld,covington2010mattr}, and repetition used a compression-based score residualized for length.

\subsubsection{Perception measures}
\label{sec:perception-measures}

After Phase-2, participants rated their LLM on seven measures covering interaction, ability, and humanness \cite{wei2023bot}. Single 1--7 items adapted from \citet{cheng2026sycophantic} measured response quality, whether LLMs behavior in this situation was right or wrong, and willingness to reuse it for similar questions. Godspeed likeability and perceived intelligence scales \cite{bartneck2009godspeed} each averaged five 1--5 semantic differential pairs. Performance and moral trust used the Multi-Dimensional Measure of Trust \cite{ullman2019trust}, each rated 0--7. Table \ref{tab:perception} reports Cronbach's $\alpha$ for every multi-item measure.

All 200 participants completed these measures, including \arm{Baseline}, in which participants rated the AI assistant's response they had read. RQ3 asks whether the three providers differed and whether ratings tracked the Phase-2 shift, so analyses use the 150 conversational participants; \arm{Baseline}'s ratings are reported descriptively alongside them (\S\ref{res:rq3}).

\subsubsection{Intake}
\label{sec:intake-measures}

The intake questionnaire describes the sample. It recorded age group, gender, race and ethnicity, education, whether English was the participant's primary language, and AI-tool familiarity and use frequency (each on a 1--5 scale). Participants also completed the Short Schwartz Value Survey (SSVS) once, rating ten values from 0 to 8 \cite{lindeman2005ssvs}.

\subsection{Procedure}
\label{sec:procedure}

\begin{figure*}[ht]
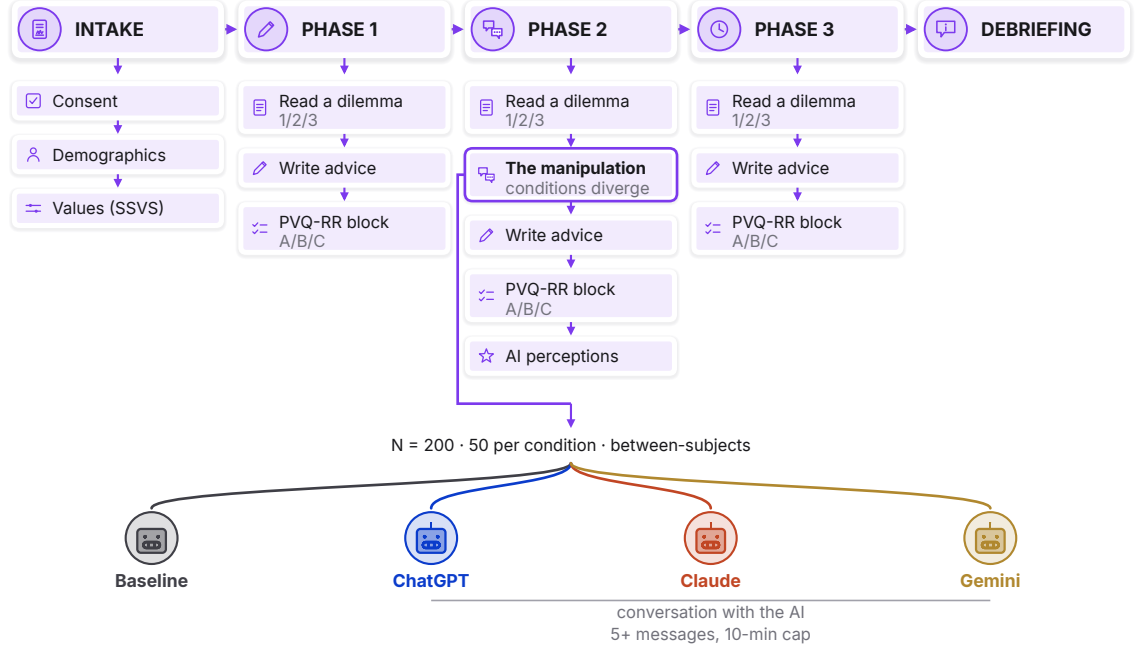

\centering
\figincl{f01_procedure_cards}
\caption{\textbf{Study procedure. Conditions differ only at Phase-2, step 2.} The three conversational arms talk with a live model for at least five messages under a ten-minute cap; \arm{Baseline} reads a fixed response to the same dilemma, synthesized from outputs of the same three models, from the same AI assistant on the same screen with the composer removed. Dilemma order and PVQ-RR block order are Williams-balanced.}
\Description{A row of five stage cards: intake, Phase-1, Phase-2, Phase-3, and debriefing. Intake lists consent, demographics, and the Short Schwartz Value Survey (SSVS). Phase-1 lists read a dilemma (1/2/3), write advice, and a Portrait Values Questionnaire--Revised (PVQ-RR) block (A/B/C). Phase-2 lists a dilemma, the manipulation where conditions diverge, write advice, a PVQ-RR block, and artificial-intelligence perceptions. Phase-3 lists read a dilemma, write advice, and a PVQ-RR block. Debriefing is a single closing card. Below Phase-2, the Baseline branch reads a fixed artificial-intelligence response synthesized from the three models' outputs, with no reply possible. The ChatGPT, Claude, and Gemini branches share the gloss conversation with the artificial intelligence, with a five- or more-message limit and a 10-minute cap. A band at the bottom lists what was held constant across all four conditions: three dilemmas, one per phase; dilemma order, counterbalanced; PVQ-RR blocks, counterbalanced; and the three phases. A sample line reads N = 200, 50 per condition, between-subjects.}
\label{fig:procedure}
\end{figure*}

Participants joined through Prolific and completed a fixed-order session in a custom application (Figure~\ref{fig:procedure}). We measured priorities before manipulation, immediately after, and after a further task, each with a different dilemma assigned by the square in \S\ref{sec:counterbalancing}.

\textbf{Intake.} After reading study information and consenting, participants completed demographics and the Short Schwartz Value Survey. They were told the study concerned advice-giving and LLM assessment, without disclosure that it measured value-priority change.

\textbf{Phase-1.} Participants read their first dilemma, wrote at least 30 words of advice, and completed a PVQ-RR block. No condition involved AI, establishing the untreated starting level for both later phases.

\textbf{Phase-2.} Participants read a second dilemma. Conversational participants talked with their assigned model while the dilemma remained visible, proceeding after five messages or the time cap. \arm{Baseline} scrolled through and read the assistant's fixed response on the same screen. All then wrote advice, completed a PVQ-RR block, and rated the AI just encountered.

\textbf{Phase-3.} To test carryover and outcome-specific durability (\S\ref{sec:rw-shifts}), participants read the remaining dilemma, wrote advice without AI in any condition, and completed the final PVQ-RR block, measuring what remains after the LLM is gone.

\textbf{Debriefing.} We disclosed the value-priority-shift objective and named the three commercial systems as examples without identifying participants' assigned model. Participants could withdraw their data, then return to Prolific with a completion code.

\subsection{Participants}
\label{sec:participants}
We recruited 200 U.S.-based adults (50 per condition) via Prolific who were fluent in English. We ran an a priori power analysis for a four-condition between-participant comparison at $\alpha = .05$. Fifty participants per condition give 85\% power to detect a medium effect (Cohen's $f = .25$) and more than 99\% power to detect a one-standard-deviation difference between any two conditions \cite{ortloff2025effectsizes}. Estimates of spread carry wider intervals than estimates of means at the same sample size, so a design that can resolve a mean shift (RQ1) can still miss a change in spread (RQ2). We therefore report a minimum detectable effect at 80\% power beside every null.

The sample included 99 women, 99 men, and 2 nonbinary participants; 57 were aged 18--29, 98 aged 30--49, 37 aged 50--64, and 8 aged 65 or older. Intake data, including SSVS scores, did not differ significantly across conditions (all $p \ge .08$; SSVS mean rating, Kruskal--Wallis $H(3) = 3.15$, $p = .370$).

Participants received \$11.96/hr; median session time was 28.8 minutes ($\mathit{IQR} = 22.0\text{--}35.3$). All 200 participants completed all three PVQ-RR blocks and entered the analysis.

\subsection{Data Analysis}
\label{sec:analysis}
We fit random-intercept mixed models by restricted maximum likelihood in \texttt{statsmodels} \cite{seabold2010statsmodels}, used \texttt{scipy} for tests, correlations, and ANOVAs, and set seed \texttt{20260805} for bootstrap, permutation, and yoking streams. We checked model convergence and fixed-effects rank. Unless noted otherwise, each outcome used
\[
\texttt{score} \sim \texttt{phase} \times \texttt{condition} + \texttt{pvq\_block} + \texttt{dilemma} + \texttt{sequence} + (1 \mid \texttt{person}),
\]
with \arm{Baseline} as the reference. A single fit supplied all phase changes and condition-minus-\arm{Baseline} contrasts, holding block, dilemma, and sequence fixed. Within-participant contrasts had 388 degrees of freedom: 600 observations minus 200 participants minus the within-person fixed-effects rank of 12.

\paragraph{Effect sizes, multiplicity, and nulls.}
We report 95\% confidence intervals and Glass's $d_{\mathrm{G}} = \widehat{\Delta}/s_{1}$ \cite{glass1976meta}, using the pooled Phase-1 standard deviation across conditions as a fixed, untreated scale; paired own-versus-yoked comparisons use $d_z$. All tests are two-sided at $\alpha=.05$, with realized-standard-error MDEs at 80\% power for null results.

Holm adjustment \cite{holm1979multiple} covered these families: three condition-minus-\arm{Baseline} contrasts per phase for the axis, spread, DSI, and advice distance; four higher-order groups per condition and contrast; three provider pairs per phase; perception measures, with pairwise tests only after a significant omnibus; and three $n$-gram lengths per condition. Own changes and Phase-2-to-Phase-3 decay contrasts are descriptive and unadjusted.

\paragraph{RQ1: value-priority shift.}
RQ1 mixed models tested priority shifts in the personal-focus axis and four centered higher-order scores, contrasting each condition with \arm{Baseline} in Phases 2 and 3 and comparing Phase-2 with Phase-3 for decay.

Because values form a circle, we also rewrite each condition's mean change $\Delta$ across the four groups as an exact change of basis, not a fitted model. Openness to Change, Self-Transcendence, Conservation, and Self-Enhancement sit at $0^{\circ}$, $90^{\circ}$, $180^{\circ}$, and
$270^{\circ}$,
\[
\Delta(\theta) = a_{0} + a\cos\theta + b\sin\theta + q\cos 2\theta.
\]
Amplitude is $\sqrt{a^{2}+b^{2}}$ and angle is $\operatorname{atan2}(b,a)$, with personal focus at $315^{\circ}$. The $q$ term is the quadrupole, the part of the change that no coordinated turn can produce. Intervals are derived from 10{,}000 bootstrap resamples of participants within each condition, with bearings averaged as unit vectors so that intervals do not wrap at $0^{\circ}$. We compared the largest gap between the three providers' bearings with the distribution of that gap under permuted provider labels, which asks whether the observed agreement is what three providers pushing in one direction would produce. We also re-estimated the Phase-2 contrast while adjusting for each participant's Phase-1 score and tested whether Phase-1 means differed across conditions.

\paragraph{RQ2: similarity of profiles.}
At each phase, to test for homogenization, we scored the leave-one-out distance from each participant's four-value profile $\mathbf{x}_{i}$ to their condition's mean with that person omitted,
\[
d_{i} = \lVert \mathbf{x}_{i} - \bar{\mathbf{x}}_{(-i)} \rVert.
\]
We entered these distances into the same mixed model, adapting the leave-one-out diversity statistic used for AI-assisted texts to value profiles \cite{doshi2024generative,anderson2024homogenization}.

To distinguish directional convergence from changes in the spread of profile magnitudes, we repeated the spread calculation separately for profile direction (unit vectors) and magnitude (Euclidean length). For this decomposition, we excluded 12 participants with zero-magnitude profiles in at least one phase, for which direction is undefined, and used the remaining 188 for all components. To account for dependence through shared centroids, permutation tests reshuffled condition labels and rebuilt all centroids \cite{anderson2006permdisp,anderson2013permanova}. Appendix~\ref{app:supporting} reports six spread definitions, leave-one-participant-out estimate ranges, and checks adjusting for Phase-1 spread or individual MRAT change.

\paragraph{RQ3: perceptions.}
RQ3 analyzed provider differences and associations with the Phase-2 shift in the 150 conversational participants. We compared providers on seven measures using one-way analyses of variance, Holm-adjusted p-values, and Welch pairwise tests after significant omnibus tests. We report \arm{Baseline} ratings descriptively in a four-condition comparison. Because the measures intercorrelate, we also tested a favorability composite averaging their $z$-scores. We regressed the signed Phase-2 axis shift on the seven measures and computed partial correlations after residualizing both sides on provider. We did not use distance between a participant's Phase-1 and Phase-3 value vectors as shift magnitude: these vectors use different parallel forms, so a norm would rectify form error rather than average it out.

\paragraph{RQ4: traces in the advice texts.}
For Phase-2 advice, written immediately after the AI exposure, uptake was the own-minus-yoked overlap (\S\ref{sec:advice-measures}),
\[
\Delta_i \;=\; m\bigl(a_i,\, s_i\bigr) \;-\; m\bigl(a_i,\, s_{y(i)}\bigr),
\]
where $m$ is either overlap measure and $y(i)$ the yoked participant. Matching sources on provider and dilemma isolates resemblance specific to the participant's exchange. One-sample $t$ tests compared $\Delta_i$ with zero; follow-up semantic tests replaced $\mathbf{s}_i$ with $\mathbf{u}_i$ to assess uptake beyond shared source content. The RQ1 mixed model also tested DSI, sentence count, vocabulary range, length-residualized repetition, and leave-one-out advice-embedding distance. Permutation tests assessed whether dilemma or condition structured the first two embedding principal components; inference applies to that two-dimensional map, not the full embedding space.

\paragraph{Manipulation checks.}
We compared conversational conditions on advice length, total words typed excluding the scripted opener, and words per typed turn. Because counts are right-skewed, we tested log words with Welch tests and Holm adjustment across three provider pairs. We compared LLM verbosity likewise, including first-reply length after the identical opener. To account for provider verbosity differences, we added words received as a covariate on the Phase-2 axis and tested provider differences with and without it. Within-provider uniformity was mean pairwise cosine across different participants' assistant turns within each dilemma, compared with the corresponding quantity across participants' own messages.

\section{Results}
\label{sec:results}

The primary value-shift and whole-profile spread analyses included all 200 participants who completed the three PVQ-RR blocks (50 per condition). We implemented the Williams-balanced design for counterbalancing \cite{Wang_Wang_Gong_2009}. At each phase, the four conditions had identical scenario sequences and PVQ-RR block compositions. Internal consistency was high for most perception measures (Table \ref{tab:perception}).

In this section, we present the results of our four research questions. We first establish that participants engaged similarly with the three LLMs, and that their visible differences do not confound the effects. We then present RQ1, which examines how value priorities changed before and after the LLM conversations. After that, in RQ2, we examine whether participants' value profiles became less dispersed in direction or magnitude after the LLM conversation. Then, in RQ3, we present whether participants' perception measures of the three LLMs are related to their value shifts. Finally, we move to RQ4, presenting whether the LLMs left their traces in the participants' written advice. In each research question, we present our statistical tests and robustness analyses. 

Unless noted otherwise, all estimates come from the mixed models in \S\ref{sec:analysis}. We use \arm{Baseline} as the reference condition and apply Holm correction within each family of tests. For null results, we report the minimum detectable effect (MDE) at 80\% power. We do not treat a nonsignificant result as evidence of no effect. Appendix \ref{app:supporting} reports the supporting analyses. Detailed results are included in the supplementary materials.

\subsection{Manipulation check}
\label{res:manipulation}

Participants engaged with all three LLMs at similar levels. Advice length, total words typed, and words per turn were comparable across conditions (smallest Holm $p = .585$; minimum detectable ratio ${\approx}1.4\times$), and each condition had a median of five turns. Each LLM also gave similar replies across participants. Replies from the same LLM had mean pairwise cosine similarities of $0.882$--$0.895$, compared with $0.492$--$0.562$ among participants' messages.

The LLMs differed in first-reply length (medians: 191 words for \arm{Claude},
151 for \arm{ChatGPT}, and 87 for \arm{Gemini}; $F(2, 147) = 267.47$, $p < .001$). Words received did not predict the Phase-2 shift ($-0.030$ per 100 words; minimum detectable slope $0.098$). Adjusting for words received did not change the comparison among the three providers (Table \ref{tab:checks}).

Time on task differed only at Phase-2. Phase-1 durations of $8.1$--$9.6$ minutes and Phase-3 durations of $5.9$--$6.5$ minutes were comparable across conditions ($H(3) = 1.37$, $p = .714$; $H(3) = 1.63$, $p = .652$). At Phase-2, the conversational conditions took $12.4$--$13.7$ minutes compared with $9.1$ minutes for \arm{Baseline}, a median gap of $3.7$ minutes ($H(3) = 20.11$, $p < .001$). Because the conversation required at least five messages, the design does not separate interaction from the additional time it required.

\subsection{RQ1: LLM conversations temporarily shifted value priorities toward personal focus}
\label{res:rq1}

\begin{figure*}[ht]
\centering
\figincl[scale=0.7]{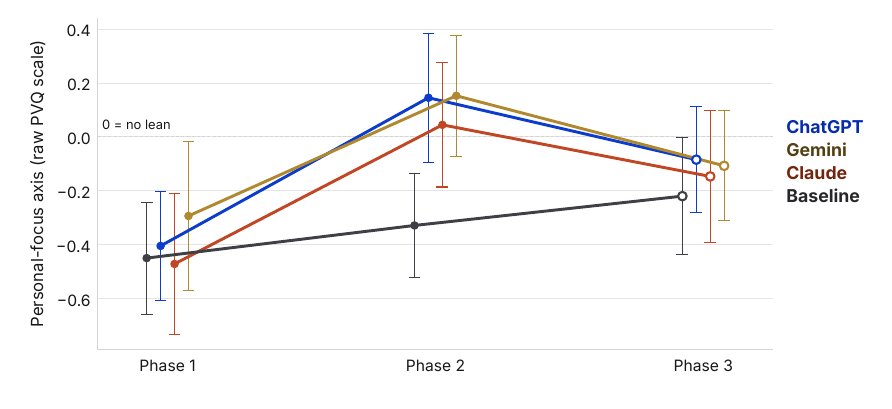}
\caption{\textbf{Reported value priorities moved toward personal focus following the exchange and moved back by the next task.} Mean personal-focus score at each phase, in raw PVQ scale points, with 95\% confidence intervals; $n = 50$ per condition. Each score is measured after that phase's dilemma and advice task. The three LLM conditions rise sharply at Phase-2 and fall by Phase-3. \arm{Baseline} rises gradually. Zero indicates equal weight on personal and social focus.}
\Description{A line chart with three phases on the horizontal axis and the personal-focus axis in raw PVQ scale points on the vertical axis. Four lines, one per condition, each with confidence interval whiskers at every phase, are labeled at their right-hand ends. All four start below zero at Phase-1. The ChatGPT, Claude, and Gemini lines rise sharply to or above zero at Phase-2, then fall back below zero at Phase-3. The Baseline line rises gradually across all three phases without a jump, ending close to where the three chat lines end.}
\label{fig:arm-trajectory}
\end{figure*}

In our study, each dilemma pitted personal interests against social interests. To highlight the value shift, we therefore used the personal focus score on the Personal Focus versus Social Focus axis as the primary outcome. The score equals Openness to Change and Self-Enhancement minus Conservation and Self-Transcendence, weighted to remain on the raw PVQ scale. All four conditions began on the social side of the Personal Focus versus Social Focus axis (Phase-1 means $-0.29$ to $-0.47$). At Phase-2, the three LLM conditions moved to the personal side. They moved back by Phase-3. \arm{Baseline} instead rose gradually across all three phases
(Figure \ref{fig:arm-trajectory}).

\begin{figure*}[t]
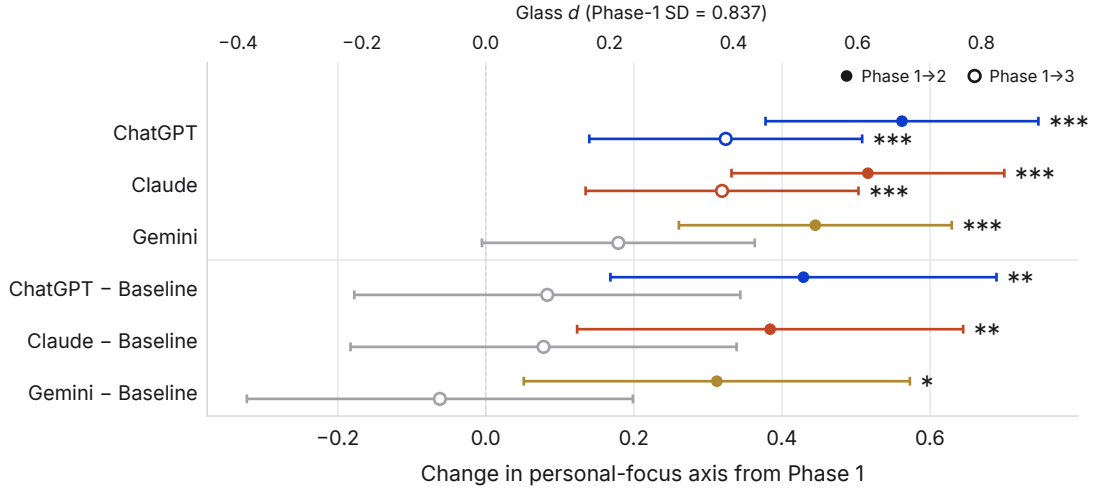

\centering
\figincl{f03_shift_forest}
\caption{\textbf{The Phase-2 value shift did not persist to Phase-3.} Change in the personal-focus axis from Phase-1, with 95\% confidence intervals. Filled dots show Phase-2, and hollow rings show Phase-3. The upper rows show each LLM condition's change; the lower rows show its contrast with \arm{Baseline}. The secondary axis expresses the same estimates as Glass $d$, using the Phase-1 SD ($0.837$). Color marks indicate estimates that survive Holm correction.}
\Description{A forest plot with six rows. The top three rows give each conversational condition's own change from Phase-1, all positive. The bottom three rows show each condition's change relative to Baseline. Each row carries two markers with confidence intervals: a filled dot for Phase-2 and a hollow ring for Phase-3. On every row, the hollow Phase-3 marker sits closer to zero than the filled Phase-2 marker. In the bottom three rows, the Phase-2 dots lie to the right of zero and are colored, while the Phase-3 rings straddle zero in gray. A secondary axis at the top gives the same estimates in Glass d units.}
\label{fig:shift-forest}
\end{figure*}

From Phase-1 to Phase-2, personal focus increased relative to \arm{Baseline} by $+0.429$ in \arm{ChatGPT} (Glass $d = 0.51$, Holm $p = .004$), $+0.384$ in \arm{Claude} ($d = 0.46$, $p = .008$), and $+0.312$ in \arm{Gemini} ($d = 0.37$, $p = .020$). The shifts were similar across providers: the largest gap was $0.145$ scale points, compared with an MDE of $0.372$ (Figure \ref{fig:shift-forest}).

By Phase-3, the LLM contrasts with \arm{Baseline} ranged from $-0.062$ to $+0.083$, and none survived correction. From Phase-2 to Phase-3, scores fell by $0.198$--$0.265$ across the LLM conditions (all $p \le .036$), returning toward their Phase-1 levels. \arm{Baseline} changed by only $+0.108$.

Self-Enhancement accounted for the shift, increasing by $+0.537$ to $+0.586$ relative to \arm{Baseline} (all Holm $p \le .002$). No condition-by-value contrast remained significant at Phase-3. Centered Self-Transcendence also fell, but the raw score changed little ($\Delta = -0.01$ to $-0.20$) while participants' mean rating across all PVQ-RR items rose ($\Delta$MRAT $= +0.07$ to $+0.26$). Subtracting this overall rise made centered Self-Transcendence appear lower; participants did not report a comparable decline in concern for others.

\begin{figure*}[t]
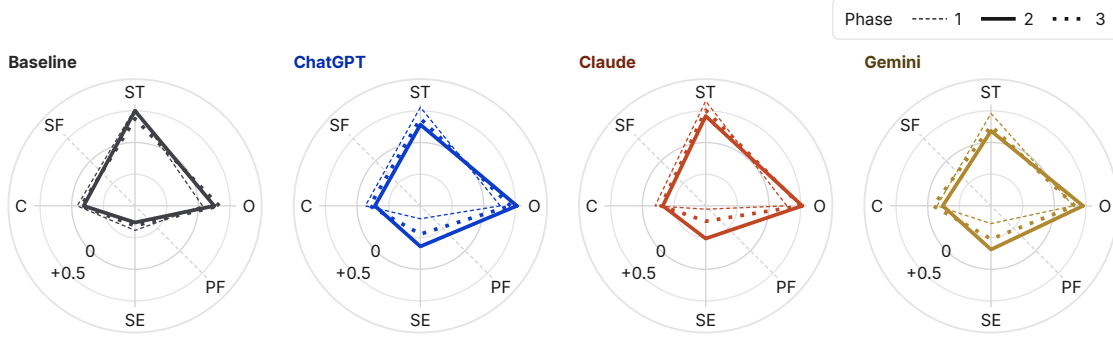

\centering
\figincl{f09_value_circle}
\caption{\textbf{The value shift was a temporary rotation toward personal focus.} Mean value profiles across the four Schwartz higher-order poles: Self-Transcendence (ST), Self-Enhancement (SE), Openness to Change (O), and Conservation (C). Radius is each value's deviation from the participant's mean rating across all 19 items; rings mark $-0.5$, $0$, and $+0.5$. Dashed, solid, and dotted outlines show Phases 1, 2, and 3. The conversational conditions shift toward the personal-focus axis in Phase-2 and return by Phase-3, whereas \arm{Baseline} drifts without returning.}
\Description{Four small radar charts in a row, one per condition, each drawn on a circle whose four compass points are the Schwartz higher-order values: Self-Transcendence at the top, Self-Enhancement at the bottom, Openness to Change at the right, and Conservation at the left. Each chart carries three overlaid quadrilateral outlines, one per phase, distinguished by line style: dashed for Phase-1, solid and heavy for Phase-2, dotted for Phase-3. In the ChatGPT, Claude and Gemini charts the solid Phase-2 outline is rotated clockwise relative to the dashed Phase-1 outline, extending toward the Self-Enhancement pole and contracting at the Self-Transcendence pole, and the dotted Phase-3 outline sits between the two. In the Baseline chart, the three outlines are displaced in the same direction rather than returning.}
\label{fig:value-circle}
\end{figure*}

Using Schwartz's value circle \cite{schwartz2012refined}, we summarized the direction and magnitude of each condition's mean change (Table \ref{tab:rotation}). At Phase-2, rotation accounted for at least $99\%$ of the change in each LLM condition. The rotations were within $15$--$25^\circ$ of the personal-focus pole, and the personal-focus axis accounted for $90$--$97\%$ of their amplitude. The three providers' bearings spanned only $10.4^\circ$, a gap that small or smaller arising in $22\%$ of permutations of the provider labels, so their directions are consistent with one common bearing; the per-condition intervals span roughly $\pm20^\circ$. \arm{Baseline} moved in a different direction with about one-third the amplitude. By Phase-3, the LLM amplitudes had fallen to $0.22$--$0.24$, indicating a weaker rotation in the same general direction (Figure \ref{fig:value-circle}).

\begin{table}[ht]
\centering
\caption{The Phase-2 change in value priorities forms one rotation. Amplitude gives the size of each condition's mean change on the value circle, and angle gives its direction. Rotation share is the percentage explained by the two rotation components. The remaining percentage is the quadrupole component, which does not describe a rotation. $0^\circ$ is Openness to Change, and $315^\circ$ is the personal-focus pole. Brackets give 95\% confidence intervals.}
\label{tab:rotation}
\small
\begin{tabular}{@{}lllr@{}}
\toprule
Condition & Amplitude & Angle & Rotation share \\
\midrule
\rowcolor{gray!15}
\multicolumn{4}{@{}l}{\textit{Phase-1 $\rightarrow$ Phase-2}} \\
\arm{Baseline} & 0.150 [0.050, 0.278] & $17^\circ$ [$325^\circ$, $67^\circ$] & 74.7\% \\
\arm{ChatGPT}  & 0.412 [0.222, 0.621] & $300^\circ$ [$280^\circ$, $326^\circ$] & 99.8\% \\
\arm{Claude}   & 0.394 [0.281, 0.522] & $298^\circ$ [$280^\circ$, $316^\circ$] & 99.1\% \\
\arm{Gemini}   & 0.363 [0.213, 0.539] & $290^\circ$ [$269^\circ$, $311^\circ$] & 99.5\% \\
\midrule
\rowcolor{gray!15}
\multicolumn{4}{@{}l}{\textit{Phase-1 $\rightarrow$ Phase-3}} \\
\arm{Baseline} & 0.168 [0.071, 0.293] & $345^\circ$ [$300^\circ$, $28^\circ$] & 54.1\% \\
\arm{ChatGPT}  & 0.240 [0.107, 0.409] & $299^\circ$ [$269^\circ$, $346^\circ$] & 99.8\% \\
\arm{Claude}   & 0.228 [0.133, 0.348] & $313^\circ$ [$279^\circ$, $345^\circ$] & 98.5\% \\
\arm{Gemini}   & 0.221 [0.094, 0.392] & $265^\circ$ [$224^\circ$, $311^\circ$] & 99.7\% \\
\bottomrule
\end{tabular}
\end{table}

The Phase-2 shift was unchanged after adjustment for Phase-1 scores (Table \ref{tab:checks}).

\subsubsection{The shift is not a property of the measurement setting}
\label{res:context}

Every PVQ-RR block followed a dilemma and an advice task, so the measurement setting itself could in principle produce a shift (\S\ref{sec:estimand}). The comparisons across conditions and phases do not support that account.

\arm{Baseline} followed the same dilemma--advice--PVQ-RR sequence and met the same AI assistant, but read rather than conversed (\S\ref{sec:control}). Its Phase-1-to-Phase-2 movement had an amplitude of $0.150$ [$0.050$, $0.278$] and a bearing of $17^\circ$, about one-third the amplitude of the conversational conditions and in a different direction from their $290$--$300^\circ$ bearings (Table \ref{tab:rotation}). The adjusted conversational contrasts remained $+0.312$ to $+0.429$. Phase-3 repeated the same measurement sequence without an LLM, but its contrasts with \arm{Baseline} ranged from $-0.062$ to $+0.083$ and none survived correction. The conversational conditions also fell $0.198$--$0.265$ from Phase-2 (all $p \le .036$). The shift therefore appeared at the manipulation phase rather than whenever the questionnaire followed a value-laden task.

The pattern was also not tied to one dilemma or a few salient value items. Rotation accounted for at least $99\%$ of the Phase-2 movement in every conversational condition. Dilemma order was counterbalanced, and the models held dilemma fixed.

Still, participants can move in the same direction without becoming more alike. RQ2 tests whether they became more similar to one another.

\begin{framed}
\begin{minipage}{\dimexpr\linewidth-10pt\relax}
\noindent\textbf{\textit{Summary.}} LLM conversations temporarily moved value priorities toward personal focus by $0.31$--$0.43$ scale points over \arm{Baseline} ($d = 0.37$--$0.51$).  Higher Self-Enhancement scores accounted for this shift. We observed the same directional pattern along the personal-versus-social axis across all three LLMs.
\end{minipage}
\end{framed}

\subsection{RQ2: Value profiles became less varied in magnitude after conversation, not more similar}
\label{res:rq2}

Participants can shift in the same direction without becoming more alike. We therefore measured the distance between each participant's value profile and the average profile for their condition to test whether they became more alike after LLM conversation. We left that participant out when calculating the condition average. Using the same mixed-model structure as RQ1, we compared changes in these distances between each LLM condition and \arm{Baseline}. 

At Phase-2, the spread contrasts with \arm{Baseline} ranged from $-0.068$ to $-0.275$, and none survived correction (smallest Holm $p = .061$). The difference appeared at Phase-3, when spread fell by $7.6\%$--$22.4\%$ in the LLM conditions but rose by $11.5\%$ in \arm{Baseline}. Relative to \arm{Baseline}, the Phase-3 contrasts were $-0.353$ for \arm{ChatGPT} (Holm $p = .006$), $-0.436$ for \arm{Gemini} ($p < .001$), and $-0.222$ for \arm{Claude} ($p = .061$). The three LLM providers did not differ at either phase.

\begin{figure*}[ht]
\centering
\figincl[scale=0.8]{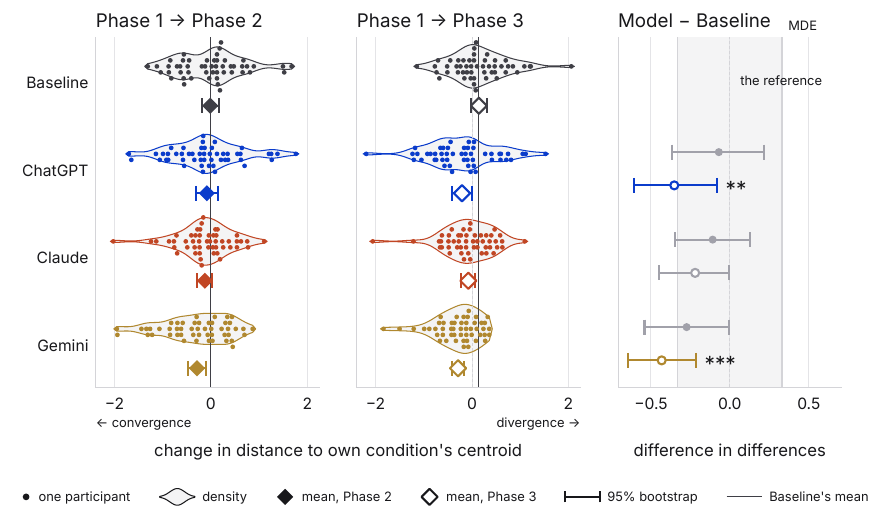}
\caption{\textbf{Between-person spread contracted only at the later endpoint.} Each dot shows one participant's change in distance from the average profile for their condition. Values to the left show less spread; values to the right show more spread. Each outline shows the distribution of the 50 values in that row, using the same bandwidth and height scale across rows. The first two panels show changes from Phase-1 to Phases 2 and 3. Diamonds show condition means with 95\% bootstrap intervals. The right-most panel compares each LLM with \arm{Baseline} and shows the MDE band ($\pm0.330$). Color marks the significance.}

\Description{A two-column dot plot. Each column holds four rows, one per condition, and each row shows fifty dots, one per participant, spread horizontally by that participant's change in distance to their condition's centroid and stacked vertically where they would collide. A pale outlined shape, the smoothed density of those same fifty values, sits behind each swarm and follows its width, so a row's shape can be read without counting dots. Below each swarm, a summary strip shows a diamond for the condition mean and a horizontal confidence interval; the diamond is filled in the Phase-2 column and hollow in the Phase-3 column. A vertical rule marks Baseline's mean in both columns. In the Phase-3 column, the three chat swarms and their means sit to the left of that rule, while Baseline's sits to the right of zero. A narrow third panel at the right plots each condition's difference from Baseline at both endpoints against a shaded MDE band; the Phase-3 markers for ChatGPT and Gemini fall outside the band.}
\label{fig:rq2-participants}
\end{figure*}

The change in spread, therefore, occurred later than the value shift: for the pooled conversational conditions, the Phase-3 contrast was $0.190$ more negative than the Phase-2 contrast ($p = .049$). Both sides contributed. Spread in
\arm{Baseline} rose by $+0.145$, whereas the spread in the LLM conditions fell by a median of
$0.150$, with decreases in all three conditions (Figure \ref{fig:rq2-participants}).

Pooling the conversational conditions, we separated directional spread from spread in profile magnitude (Table \ref{tab:decomposition}). The \textit{contraction} in spread at Phase 3 reflected a reduction in between-person variation in profile magnitude. The Phase-3 difference in changes relative to \arm{Baseline} was $-0.255$ ($p = .004$), equivalent to $39\%$ of Phase-1 spread. We detected no corresponding contraction in directional spread ($-0.038$, $p = .601$). A circular decomposition showed the same pattern. Angular spread changed by only $-5.2^\circ$ (MDE $22.8^\circ$), while amplitude spread decreased descriptively by $-0.111$.

The Phase-3 contraction was robust to randomization inference, alternative definitions of spread, adjustment for Phase-1 spread, and leave-one-participant-out analyses (Table \ref{tab:rq2-robustness}). We found no evidence that participants converged on a common value ranking. The contraction instead reflected reduced between-person variation in profile magnitude. \S\ref{res:tailoring} tests whether their written advice became more alike.

\begin{table}[ht]
\centering
\caption{Changes in spread of whole profiles, directions, and magnitudes, comparing pooled participants from \arm{ChatGPT}, \arm{Claude}, and \arm{Gemini} with \arm{Baseline} ($N = 188$). Spread is calculated within each original condition. Estimates are differences in changes from Phase-1 to each later phase; negative estimates indicate greater contraction in the conversational conditions. Percentages express Phase-3 contrasts relative to Phase-1 spread.}
\label{tab:decomposition}
\small
\begin{tabular}{@{}lrrrrr@{}}
\toprule
& \multicolumn{2}{c}{
  \colorbox{gray!15}{\strut\hspace{1.2em}P1 $\rightarrow$ P2\hspace{1.2em}}
}
& \multicolumn{2}{c}{
  \colorbox{gray!15}{\strut\hspace{1.2em}P1 $\rightarrow$ P3\hspace{1.2em}}
}
& \colorbox{gray!15}{\strut\hspace{0.5em}P3 as \% of\hspace{0.5em}} \\
\cmidrule(lr){2-3}\cmidrule(lr){4-5}
Space & Estimate & $p$ & Estimate & $p$ & P1 spread \\
\midrule
Whole profile      & $-0.127$ & .267 & $\mathbf{-0.346}$ & $\mathbf{<.001}$ & $-26.7\%$ \\
Shape only         & $+0.143$ & .037 & $-0.038$ & .601 & $-4.8\%$ \\
Magnitude only     & $-0.075$ & .446 & $\mathbf{-0.255}$ & $\mathbf{.004}$ & $-39.0\%$ \\
\bottomrule
\end{tabular}
\end{table}

\begin{framed}
\begin{minipage}{\dimexpr\linewidth-10pt\relax}
\noindent\textbf{\textit{Summary.}} By Phase-3, value profiles were less spread out relative to \arm{Baseline}. The pooled decomposition located this contraction in the spread of profile magnitudes, with no detected contraction in directional spread.
\end{minipage}
\end{framed}

\subsection{RQ3: All three LLMs were rated favorably, and we detected no provider differences}
\label{res:rq3}

Differences in how participants viewed the LLMs could explain the value shift. For example, a better-liked LLM might have been more influential. We therefore tested whether the three LLMs differed on the perception measures and whether those measures were associated with participants’ value shifts. The tests below are estimated on the 150 participants in the LLM conditions; we report \arm{Baseline} ratings descriptively at the end of this section.

\begin{figure*}[ht]
\centering
\figincl[scale=0.8]{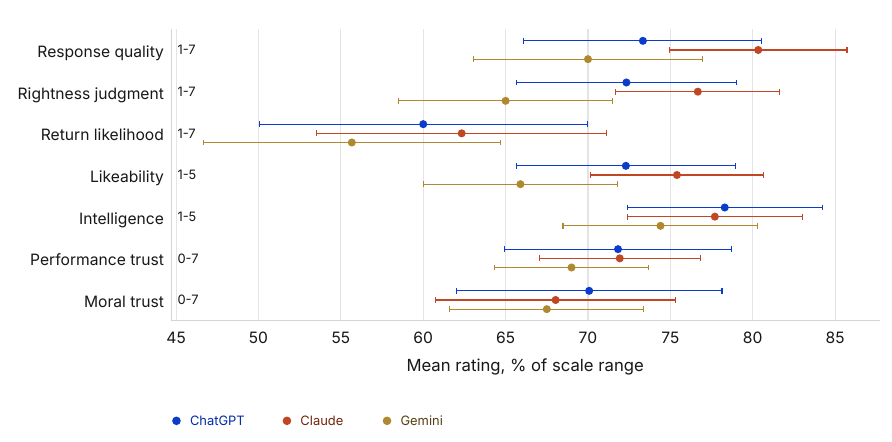}
\caption{\textbf{Participants rated all three LLMs favorably.} Means and 95\% confidence intervals are shown as percentages of each measure's scale range so that the 0--7, 1--5, and 1--7 scales can be compared. No measure differed by provider after Holm correction. The largest gap was $0.70$ scale points for rightness, compared with an MDE of $0.73$. The combined favorability score also did not differ, $F(2, 147) = 1.30$, $p = .275$.}
\Description{Seven rows, one per perception measure: response quality, rightness judgment, return likelihood, likability, intelligence, performance trust, and moral trust. Each row contains three colored dots representing confidence intervals, one per provider, positioned on a horizontal axis showing the mean rating as a percentage of that measure's scale range, and labeled with the response scale used. Every dot sits well above the fifty percent midpoint, between roughly fifty-five and eighty percent, and the three providers' intervals overlap heavily on every row. A key below names the three providers in their own colors.}
\label{fig:perception}
\end{figure*}

Every LLM received a mean rating above the midpoint of every scale, with means ranging from $56\%$ to $80\%$ of the scale range. No measure differed by provider after Holm correction; the largest observed gap was $0.70$ scale points, compared with an MDE of $0.73$ (Figure \ref{fig:perception}; Table \ref{tab:perception}).

The seven perception measures were strongly correlated (mean $r = 0.78$), so we summarized them with a standardized composite favorability score ($\alpha = 0.96$). This score also did not differ by provider, $F(2,147) = 1.30$, $p = .275$.

The seven perception measures jointly explained $7.5\%$ of variation in the Phase-2 shift, but the regression was not statistically significant ($p = .131$). No individual Pearson correlation survived Holm correction. In the separate partial-correlation analysis, adjusting for provider left the strongest association, with response quality, essentially unchanged. These ratings do not explain the shared shift, although our sample cannot rule out modest differences between providers.

In the \arm{Baseline} condition, participants also evaluated the AI favorably, with every mean above the corresponding scale midpoint. Its overall favorability score was $0.20$ SD below the pooled conversational conditions, a difference we did not detect as significant ($p = .197$; MDE $d = 0.48$). We therefore detected no perceptual advantage for interaction over the fixed-response AI-attributed \arm{Baseline}.

\begin{framed}
\begin{minipage}{\dimexpr\linewidth-10pt\relax}
\noindent\textbf{\textit{Summary.}} Participants rated all three LLMs above the midpoint of every scale, and no difference between providers survived correction. These ratings also did not predict participants' value shifts, so they do not explain the shared shift.
\end{minipage}
\end{framed}

\subsection{RQ4: Participants used language from their own LLM in their advice and wrote more semantically focused advice}
\label{res:rq4}

We next turn from what participants reported to what they produced. At Phase-2, participants wrote about the dilemma they had just discussed with an LLM. Phase-1 occurred before the conversation, and Phase-3 used a new dilemma without an LLM. We first test whether participants reused words or ideas from their own LLM's replies. We then test how their advice changed and whether participants assigned to the same provider wrote more similar advice.

\subsubsection{Participants reused their own LLM's words and meaning}
\label{res:uptake}

Two texts about the same dilemma will share some words even if neither influenced the other. We therefore compared each advice text with two sets of LLM messages. The first contained the messages that the participant's own LLM sent them. The second contained LLM messages from another participant's conversation with the same model about the same dilemma, which we call the matched (\emph{yoked}) conversation. This comparison holds the dilemma and LLM provider constant while varying whether the messages came from the participant's own interaction or a matched one. Any remaining difference, therefore, captures how much more the advice resembled the messages the participant's own LLM sent.

\begin{figure*}[ht]
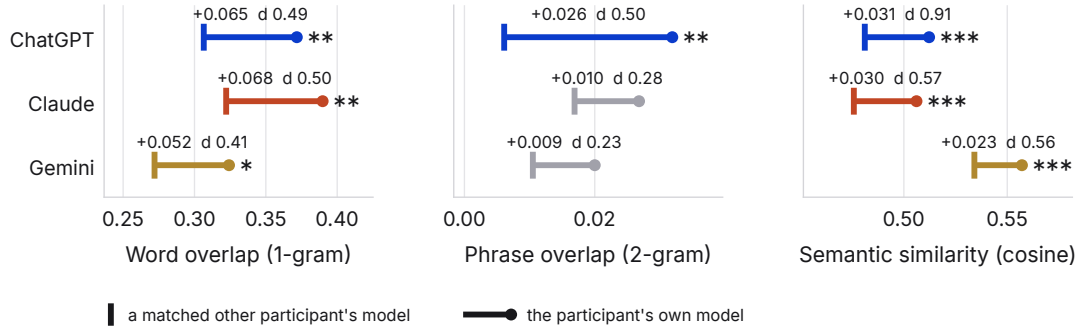

\centering
\figincl{f12_uptake}
\caption{\textbf{Participants' advice resembled their own LLM's messages.}  Panels show content-word overlap, two-word phrase overlap, and cosine similarity between the advice and source-message embeddings. Ticks show similarity to LLM messages from another participant's conversation with the same model about the same dilemma. Dots show similarity to the messages the participant's own LLM sent. All three providers showed differences in content-word overlap and semantic similarity.}

\Description{Three panels side by side, one per measurement grain: word overlap, phrase overlap, and semantic similarity. Each panel holds three rows, one per provider. Each row is a horizontal gap chart: an upright tick marks the value against a matched other participant's model, a heavy segment runs rightward from it, and a dot marks the value against the participant's own model. In every row of every panel, the dot lies to the right of the tick, so the own-model value is the higher of the two. The gaps are widest in the word-overlap panel and narrowest in the phrase-overlap panel.}
\label{fig:uptake}
\end{figure*}

Phase-2 advice shared more content words with the participant's own LLM than with the matched messages. The differences were $+0.065$ for \arm{ChatGPT}, $+0.068$ for \arm{Claude}, and $+0.052$ for \arm{Gemini} ($d_z = 0.49$, $0.50$, and $0.41$). Semantic similarity was also higher by $+0.031$, $+0.030$, and $+0.023$ ($d_z = 0.91$, $0.57$, and $0.56$; all Holm $p < .001$). Two-word phrase overlap survived correction only for \arm{ChatGPT} ($+0.026$, $p = .003$; Figure \ref{fig:uptake}). The longest shared sequence averaged only $1.4$--$1.6$ content words, indicating reuse rather than verbatim copying.

As a positive control, \arm{Baseline} advice more closely resembled the passage participants read than a passage about an unseen dilemma ($+0.134$, $d_z = 1.23$, $p < .001$).

\subsubsection{Participants' own writing became more semantically focused}
\label{res:dsi}

We next tested whether the advice became more focused. We used Divergent Semantic Integration (DSI), the mean pairwise cosine distance among the sentences in each advice text. A high DSI means that the text covers more distinct ideas. A low DSI means that it stays focused on a narrower set of ideas. We compared each participant's DSI across phases while holding the dilemma fixed. 

\begin{figure*}[ht]
\centering
\figincl[scale=.8]{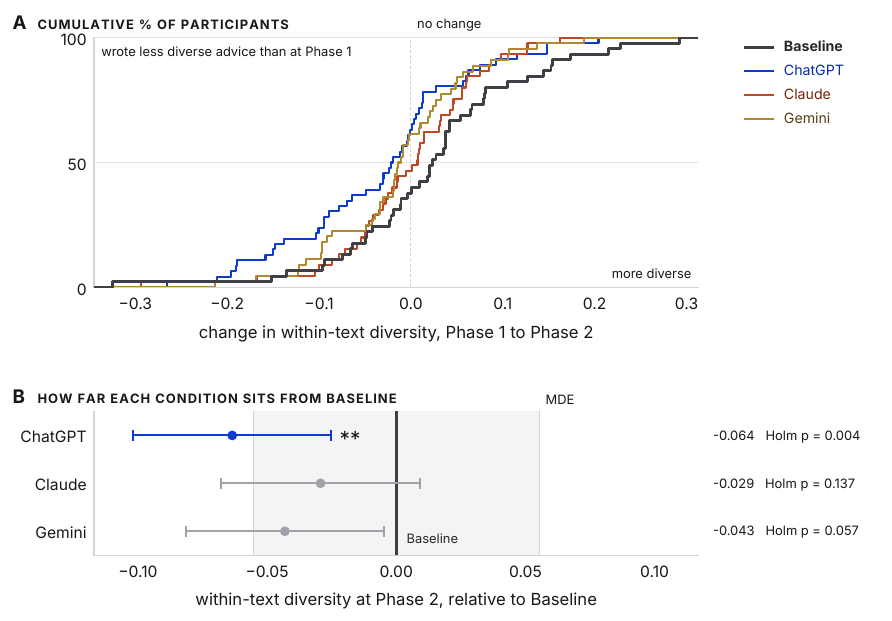}
\caption{\textbf{Advice shifted toward greater semantic focus, most clearly in \arm{ChatGPT}.} Within-text diversity is the mean pairwise cosine distance among an advice text's sentences. \textbf{(A)} Cumulative distributions of each participant's Phase-1-to-Phase-2 change. \textbf{(B)} Adjusted contrasts with \arm{Baseline}, shown against the MDE band ($0.055$). All estimates are negative; only \arm{ChatGPT} survives Holm correction ($-0.064$, $p = .004$).}
\Description{Two stacked panels sharing a horizontal axis. The top panel plots four cumulative distribution curves, one per condition, over each participant's change in within-text diversity from Phase-1 to Phase-2. The three conversational conditions' curves sit to the left of the Baseline curve and hold that order across the range, meaning more of their participants wrote less diverse advice than before. The bottom panel plots three rows, one per chat condition, each a point estimate of the difference from Baseline with a confidence interval, drawn against a heavy vertical line at zero that represents Baseline and a shaded minimum-detectable-difference band. All three estimates are negative; ChatGPT's and Gemini's intervals lie wholly left of zero, and Claude's crosses it.}
\label{fig:dsi-change}
\end{figure*}

Relative to \arm{Baseline}, the adjusted Phase-2 DSI contrast was $-0.064$ for \arm{ChatGPT} (Holm $p = .004$), $-0.043$ for \arm{Gemini} ($p = .057$), and $-0.029$ for \arm{Claude} ($p = .137$). All three estimates favored greater semantic focus, but only the \arm{ChatGPT} contrast survived correction (Figure \ref{fig:dsi-change}).

We found no corresponding changes in vocabulary, repetition, sentence count, or advice length.


The change in DSI was also temporary. By Phase-3, each LLM condition was within $0.003$ of its Phase-1 level, and none differed from \arm{Baseline} (MDE $0.055$). Thus, semantic focus followed the same Phase-2-to-Phase-3 pattern as the value shift.

\subsubsection{The influence was tailored to each conversation, and the advice did not converge}
\label{res:tailoring}

Participants can draw from their own conversations without becoming more alike as a group. To separate individualized influence from global influence, we split each participant's LLM messages into components shared with other participants in the same condition and dilemma, and the remainder specific to that participant. We then repeated the comparison between each participant's own conversation and another participant's conversation with the same model about the same dilemma (\S\ref{res:uptake}). Advice was more similar to the participant-specific part of their own LLM's messages than to that of the matched messages. The differences were $+0.099$, $+0.095$, and $+0.066$ for \arm{ChatGPT}, \arm{Claude}, and \arm{Gemini} ($d_z = 0.92$, $0.60$, and $0.56$).

\begin{figure*}[ht]
\centering
\figincl[scale=0.8]{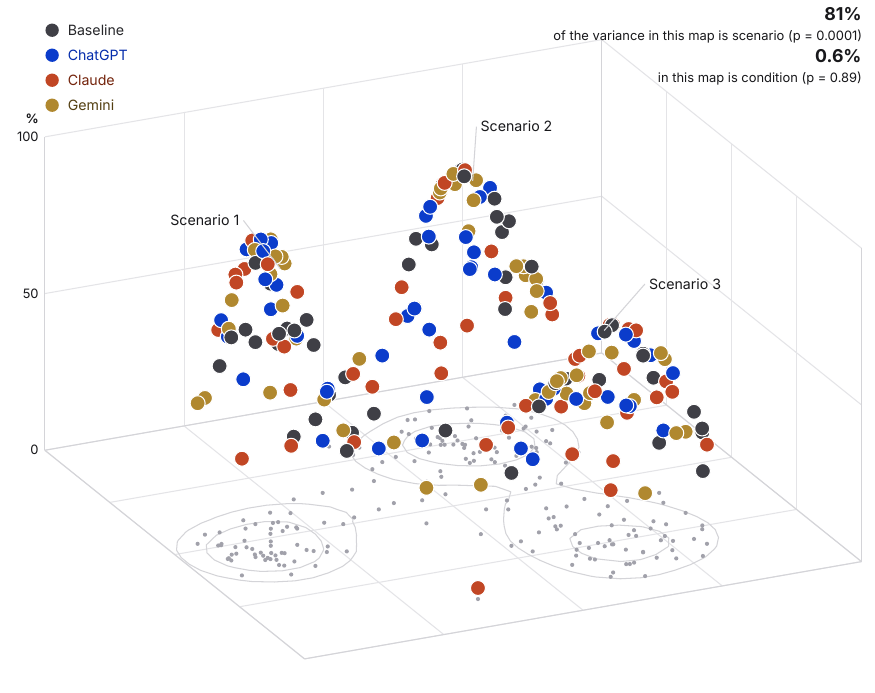}
\caption{\textbf{Advice grouped by dilemma, not by provider} All 200 Phase-2 advice texts in a single embedding landscape. Height is the density of the pooled sample, and color is the condition, interleaved throughout. Three islands appear, one per scenario: $81\%$ of the variance in this map is scenario ($p = .0001$) and $0.6\%$ is condition ($p = .89$).}
\Description{A three-dimensional density landscape over a two-dimensional embedding of all two hundred Phase-2 advice texts, drawn in orthographic projection on an isometric floor. The surface forms three distinct hills, each labeled with one of the three advice scenarios. Individual texts are drawn as small points colored by condition, and each hill contains points of all four colors mixed together, rather than any hill belonging to one condition. A key states that height is density as a percentage.}
\label{fig:advice-pooled}
\end{figure*}

The dilemma itself strongly shaped the advice. Among all 200 Phase-2 texts, the dilemma explained $81\%$ of variation in the two-dimensional embedding map ($p < .001$), while condition explained $0.6\%$ ($p = .890$). The four conditions were mixed within the three dilemma groups (Figure \ref{fig:advice-pooled}). Pairwise comparisons and comparisons across phases showed the same pattern, so all convergence tests controlled for dilemma.

We detected no condition-level convergence in advice. Relative to \arm{Baseline}, changes in leave-one-out cosine distance ranged from $-0.0009$ to $+0.0167$ (all Holm $p = 1.00$). The providers also produced similarly dispersed advice. At the participant level, uptake was not related to convergence in advice or value profiles. The exchange, therefore, left detectable traces in individual advice texts without detectable convergence across participants.

\begin{framed}
\begin{minipage}{\dimexpr\linewidth-10pt\relax}
\noindent\textbf{\textit{Summary.}} Participants aligned locally with their own LLM conversation rather than globally with one another. They reused their model's words and meaning in advice they wrote themselves. This influence was personalized through LLM replies shaped by their own questions. Their advice also shifted toward greater semantic focus as a trend without becoming shorter, more repetitive, or less lexically diverse, and returned to its Phase-1 level on the same timeline as the value shift.
\end{minipage}
\end{framed}

\section{Discussion}

Our results reveal two important patterns. First, brief interactions with each of three commercial LLMs produced the same temporary reorientation of participants' reported values toward a more personal focus, with higher relative priorities assigned for Openness to Change and Self-Enhancement, over Conservation and Self-Transcendence. Higher Self-Enhancement accounted for most of this shift, while raw Self-Transcendence changed little. The movement formed a coherent rotation around Schwartz's value circle, appeared only during the LLM phase, and had receded by the next task. 

Second, this reorientation occurred without detectable convergence in value directions or advice. Their advice retained words and meaning from their own exchanges and became more semantically focused. The later contraction in value-profile spread in Phase-3 instead reflected reduced between-person variation in profile magnitude. Participants also rated all three LLMs favorably, but these perceptions neither distinguished the providers nor explained the shift.

We begin by interpreting the effects on participants' value priorities as a temporary reorientation that occurs during the period in which judgment was being formed. We then turn to its direction. Personal focus is culturally situated and corresponds to the broader WEIRD-aligned orientation documented in prior audits of commercial LLMs. We consider what it means that three systems moved participants along this shared axis through locally different exchanges, and how the common movement can occur without detectable convergence in value directions or advice. Together, these findings raise implications for how HCI understands LLMs' impacts on the homogenization of outputs users produce and on their values, and evaluates the neutrality of non-directive LLM decision support.

\subsection{A Temporary Reorientation of Value Priorities}

A conversation lasting no more than ten minutes shifted participants' reported value priorities toward personal focus by $0.31$--$0.43$ scale points over the non-interactive AI control ($d=0.37$--$0.51$). The pattern appeared independently in the \arm{ChatGPT}, \arm{Claude}, and \arm{Gemini} conditions, under the same thinking-partner prompt and opening question. By Phase-3, the contrasts with \arm{Baseline} had shrunk to between $-0.062$ and $+0.083$, and none survived correction. The mean profiles in all three LLM conditions, therefore, moved around Schwartz's value circle in the same direction and had returned toward their earlier positions by the next task.

Value theory provides a direct way to interpret this timing. A person's broader value hierarchy generally changes slowly, through sustained experience and identification \cite{bardi2009structure, bardi2011dual, kekes1993}. The values that become active during a particular judgment are more responsive to context, and these activated values are especially predictive of the choice made at that moment \cite{verplanken2002motivated, sagiv2017values}. Research on construct accessibility similarly shows that recently activated ideas can organize subsequent judgment before their influence weakens with time \cite{higgins1977category, srull1979accessibility, higgins1985priming}. The Phase-2 movement fits this shorter process: the conversation temporarily changed the relative prominence of the values as participants reasoned and responded.

The timing of this effect is practically significant, as values active during a judgment are especially likely to organize that judgment \cite{verplanken2002motivated, sagiv2017values}, and decision-support interactions often occur while users are forming a recommendation or deciding what to do \cite{chatterji2025}. A shift concentrated in this interval can therefore shape reasoning even when it recedes before a later task. This implication also matters for systems that elicit values through an LLM. A profile collected immediately after an exchange may partly reflect priorities made salient by the elicitation setting itself. Pre-interaction measurement, independent elicitation, or delayed reassessment would help distinguish a user's broader priorities from those made prominent during the exchange.

One way to picture the result is as a compass briefly deflected by a nearby magnetic field: its reading turns under a local influence and relaxes once that influence passes. The analogy captures both the geometry and timing of the result. Rotation accounted for at least 99\% of the Phase-2 movement in every LLM condition, meaning that the four higher-order values turned together around Schwartz's circle. \arm{Baseline} participants completed the same measurement sequence and encountered the same AI framing, yet showed a smaller, gradual movement on a different bearing. This image connects to \citet{riedl2026}, who describe interaction with LLMs as a ``social forcefield'' whose effects extend into subsequent interactions with other people. The field observed here followed a shorter trajectory: its strongest effect was concentrated around the LLM interaction and had weakened by the next task. 

This shorter trajectory adds a different timescale to the growing literature on carryover from human--AI interaction. Previous work has found that LLM-assisted writing can shift subsequent attitudes \cite{jakesch2023cowriting,Williams-Ceci_Jakesch_Bhat_Kadoma_Zalmanson_Naaman_2026}, that users' confidence can remain aligned with an AI advisor in later unaided decisions \cite{li2025confidence}, and that conversational effects can extend to self-perception and later interactions with other people \cite{li2026personality,riedl2026,harrell2025}. Other effects, including changes in conspiracy beliefs, have persisted over considerably longer periods \cite{costello2024conspiracy}. 

Taken together, these findings suggest that the duration of an LLM's influence depends on what is being shifted and how the interaction activates it. An AI's social influence can extend beyond the immediate exchange on different timescales, sometimes persisting into later interactions and sometimes remaining concentrated within the period in which judgment is being formed. In our study, value priorities changed sharply within the conversational context and then receded rapidly.

\begin{quote}
\textbf{Implication 1: LLM influence can be temporally concentrated.} The value shift was strongest immediately after interaction and had receded by the next task, positioning the effect within the period in which participants were forming their judgments.
\end{quote}

Beyond its temporal aspect, the magnitude of this immediate movement is striking. Experimental attempts to change reported value priorities have generally used interventions that explicitly name, prime, or challenge particular values, producing small-to-medium effects \cite{maio1998truisms, maio2009changing, arieli2014benevolence, russo2022values}. Our prompt named no values, assigned no persuasive direction, and explicitly instructed the LLM to leave the recommendation to the participant. Nevertheless, the observed effects fell within the same broad range. The comparison with \arm{Baseline} further locates this movement beyond exposure to AI-attributed considerations alone, although the present design cannot separate the back-and-forth exchange from the additional reflection and time it occasioned.

\begin{quote}
\textbf{Implication 2: Value activation does not necessarily require value-explicit prompting.} A thinking-partner interaction without naming values and any persuasive direction still produced a coordinated reorientation comparable in size to deliberate value interventions.
\end{quote}

In summary, across all three systems, the interaction produced a short-lived but directionally consistent reweighting of the reported priorities to judgment without an explicit mention of values. Why, then, did three different LLMs all turn participants toward personal focus?

\subsection{The Same Direction, Different Paths}

The most plausible place to begin is with LLMs' existing orientations. Across value frameworks and evaluation settings, mainstream systems have repeatedly expressed priorities aligned with WEIRD populations \cite{birhane2022,kazemi2024,tao2024,cao2023,zhou2025,sambasivan2021, rahman2026ccd}. This pattern has appeared when models were administered the Schwartz PVQ-RR directly \cite{hadar2024}, and some value emphases persisted across paraphrases, translations, and assigned personas \cite{moore2024valueconsistency,lee2026valueinertia}. Personal focus occupies the same broad cultural direction: national samples differ systematically along this axis, with WEIRD populations placing greater relative weight on autonomy, achievement, and the expression of personal interests \cite{schwartz2004,cao2023,henrich2010}. The turn observed here, therefore, connects two bodies of evidence that have largely been studied separately. Prior work locates a value skew in what commercial LLMs produce; our results locate a corresponding movement in the people interacting with them.

The same direction across three model families points toward a regularity broader than the disposition of one provider. Commercial systems differ in architecture, presentation, and style, but they are developed within overlapping linguistic, cultural, and alignment environments \cite{birhane2022}. Prior work has warned that these environments can narrow the range of values represented by aligned systems \cite{sorensen2024pluralistic} and contribute to an algorithmic monoculture even when individual outputs appear varied \cite{kleinberg2021}. Our findings suggest that such commonality may also be visible in the direction of LLMs' influence on users' values.

That reorientation appeared even among participants from the United States, one of the populations already most closely associated with this cultural profile. All four condition means began on the social side of the personal-social axis, yet the three LLM conditions crossed toward personal focus at Phase-2.

\begin{quote}
\textbf{Implication 3: A shift toward personal focus can emerge as a shared cross-provider impact of LLM use.} Three commercial LLMs produced the same broad change along a culturally situated value axis, connecting prior model-side evidence of WEIRD value skew with a corresponding human side shift resulting from interaction.
\end{quote}

The common movement occurred without detectable convergence in value directions or advice. The estimates instead describe a shared vector of movement with different starting and ending points. Participants moved toward personal focus without detectable convergence in value directions. This distinction complicates existing accounts of homogenization, which have primarily examined whether AI-assisted outputs become more alike \cite{agarwal2025, doshi2024generative, veselovsky2025}. Our results indicate that a systematic movement in users' reported value priorities can occur without detectable convergence in their advice.

The RQ4 results help explain how these two patterns can coexist. Participants reused more words and meanings from their own exchange than from another participant's conversation with the same model about the same dilemma (the matched exchange).  Interactive alignment provides one account of how language becomes coordinated within such an exchange \cite{pickering2004dialogue, brennan1996}. Participants then reproduced parts of that language in advice they authored themselves, a process that prior work connects to subsequent judgments through saying-is-believing and self-perception effects \cite{higgins1978saying, bem1972selfperception}. Uptake did not statistically explain individual value shifts, so it does not establish the mechanism behind them. It does, however, show that each local exchange entered the participant's subsequent reasoning. Different conversational paths could therefore accompany the same aggregate shift without detectable convergence in advice.

The textual changes followed this pattern. Advice in the LLM conditions became more semantically focused relative to \arm{Baseline}, most clearly in the \arm{ChatGPT} condition, without becoming shorter, more repetitive, less lexically diverse, or detectably more similar across participants. The change was concentrated within individual responses rather than expressed as standardization across the group. At Phase-3, the value results showed another form of collective change: between-person spread in profile magnitudes contracted, with no detected contraction in directional spread. Mean profiles had moved back toward their Phase-1 positions, while between-person variation in profile magnitude had narrowed. An average return can therefore coexist with a change in the distribution around that average.

\begin{quote}
\textbf{Implication 4: Directional value influence and value and output homogenization are empirically distinct.} Participants moved in the same average direction without detectable convergence in value directions or advice.
\end{quote}

These distinctions matter for how HCI studies evaluate AI's influence on collective diversity \cite{doshi2024generative,anderson2024homogenization}. Measuring only the similarity of outputs would have missed the shared movement in value priorities. Measuring only the mean shift would have missed the later contraction in the spread of profile magnitudes. Claims about homogenization should therefore specify what has become more alike: the content people produce, the direction of their value priorities, or the magnitude with which those priorities are expressed. These outcomes diverged empirically in our study.

The findings also complicate familiar ideas of neutrality and user control. The thinking-partner prompt withheld recommendations, named no values, and repeatedly left the decision with the participant. Participants in the LLM conditions nevertheless shifted in the same direction. Neutrality at the level of verdicts, therefore, leaves open a deeper question of which considerations an LLM makes salient. Participants' favorable ratings offered little indication of this movement: the three providers were evaluated similarly, and those evaluations did not predict who shifted.

\begin{quote}
\textbf{Implication 5: Non-directive support can still be normatively directional.} The LLMs withheld recommendations and left the decision to participants, yet the mean value profiles shifted in a common direction that participants' favorable evaluations did not reveal.
\end{quote}

Overall, the findings describe a form of influence subtler than copying, overt persuasion, or uniform output. Three LLMs supported individualized exchanges that left distinct traces in participants' advice, while the mean profiles in all three conditions moved along the same value axis, even though dilemmas were not personal. Research on sycophancy in personal conflicts and emerging reports of AI-associated delusions highlight concerns about reinforcement when users are personally involved \cite{cheng2026sycophantic,morrin2026delusions}. What happens when the dilemma is one's own? Greater personal relevance could invite longer exchanges, deeper disclosure, and repeated engagement, potentially leaving a larger or more persistent footprint in users' value priorities and in how they reason. 


\subsection{Limitations}

This study examined 200 adults in the United States using three real dilemmas shared by people seeking advice, each selected because it placed personal and social values in tension. The consistency of the Phase-2 shift across all three LLM conditions strengthens the finding within this sample. Personal stakes were neither manipulated nor measured, so the study cannot estimate whether influence would be stronger in personally consequential situations. Replications across cultures, domains, value dimensions, and interaction roles can further solidify the result while exploring where its magnitude or direction varies.

The non-interactive AI control retained the dilemma, interface, AI attribution, and exposure to AI-generated considerations while removing the exchange. Live interaction also required more time and active engagement. The study, therefore, estimates the overall effect of the interaction. Time-matched controls and designs that vary the back-and-forth exchange, participant elaboration, and time on task independently would help explain how the shift arises. Repeated administration of the PVQ-RR may also have heightened attention to values, although the common sequence across conditions limits this as an explanation for the treatment--control difference.

The PVQ-RR captures priorities as participants reported them at three points within a single session. Phase-3 shows that the shift had receded by the next task, while leaving its precise duration and behavioral consequences open. Longer follow-ups, behavioral measures, and repeated-use studies can trace whether similar shifts recur, diminish with familiarity, or accumulate over time.

Finally, testing three leading systems reduces dependence on any single provider, although the findings remain tied to the model versions, shared prompt, and chat interaction examined here. Replication across model updates and interaction designs can show which features of the result endure. Larger samples would also provide greater precision for estimating small provider differences and changes in between-person spread.

\section{Conclusion}

This work investigated whether interacting with an LLM can temporarily shift the values people prioritize while reasoning about a decision. In a preregistered study with 200 U.S. adults, participants in the \arm{ChatGPT}, \arm{Claude}, and \arm{Gemini} conditions moved toward greater personal focus relative to \arm{Baseline}, the non-interactive AI control. This pattern emerged across all three systems, even though the thinking-partner prompt did not mention values or instruct the LLMs to favor a particular direction. The shift had receded by the next task. Participants also carried words and meaning from their own exchanges into the advice they subsequently wrote, without detectable convergence in value directions or advice.

These findings show that LLM interaction can shape the values active during a judgment without producing uniform or durable change. This is important for decision-support systems because even a temporary shift may coincide with the period in which a recommendation or decision is being formed. More broadly, the effects of LLM use extend beyond what these systems produce to include changes in the people interacting with them, including what they prioritize and how long those changes remain.

\bibliographystyle{ACM-Reference-Format}
\bibliography{references}

@article{lindeman2005ssvs,
  author  = {Lindeman, Marjaana and Verkasalo, Markku},
  title   = {Measuring Values With the Short Schwartz's Value Survey},
  journal = {Journal of Personality Assessment},
  year    = {2005},
  volume  = {85},
  number  = {2},
  pages   = {170--178},
  doi     = {10.1207/s15327752jpa8502_09}
}

@article{Wang_Wang_Gong_2009, title={The Construction of a Williams Design and Randomization in Cross-Over Clinical Trials Using SAS}, volume={29}, rights={Copyright (c) 2008 Bing-Shun Wang, Xiao-Jin Wang, Li-Kun Gong}, ISSN={1548-7660}, DOI={10.18637/jss.v029.c01}, abstractNote={A Williams design is a special and useful type of cross-over design. Balance is achieved by using only one particular Latin square if there are even numbers of treatments, and by using only two appropriate squares if there are odd numbers of treatments. PROC PLAN of SAS/STAT is a practical tool, not only for random construction of the Williams square, but also for randomly assigning treatment sequences to the subjects, which makes integration of the two procedures possible. The present paper provides a general SAS program for the random construction of a Williams design and the relevant procedure for randomization. Examples of a three-treatment, three-period (3 x 3) and a four-treatment, four-period (4 x 4) cross-over designs are given to illustrate the function of the SAS program. The results can be regenerated and replicated with the same random number seed. The general SAS program meets the practical needs of researchers in the application of Williams designs.}, journal={Journal of Statistical Software}, author={Wang, Bing-Shun and Wang, Xiao-Jin and Gong, Li-Kun}, year={2009}, month=feb, pages={1--10}, language={en},
  number = {Code Snippet 1}
}

@inproceedings{sorensen2024pluralistic,
  author    = {Sorensen, Taylor and Moore, Jared and Fisher, Jillian and
               Gordon, Mitchell L. and Mireshghallah, Niloofar and
               Rytting, Christopher Michael and Ye, Andre and Jiang, Liwei and
               Lu, Ximing and Dziri, Nouha and Althoff, Tim and Choi, Yejin},
  title     = {Position: A Roadmap to Pluralistic Alignment},
  booktitle = {Proceedings of the 41st International Conference on Machine Learning},
  series    = {Proceedings of Machine Learning Research},
  volume    = {235},
  pages     = {46280--46302},
  year      = {2024},
  publisher = {PMLR},
  address   = {Vienna, Austria},
  url       = {https://proceedings.mlr.press/v235/sorensen24a.html}
}

@inproceedings{kirk2024prism,
  author    = {Kirk, Hannah Rose and Whitefield, Alexander and R{\"o}ttger, Paul and
               Bean, Andrew and Margatina, Katerina and Ciro, Juan and
               Mosquera, Rafael and Bartolo, Max and Williams, Adina and He, He and
               Vidgen, Bertie and Hale, Scott A.},
  title     = {The {PRISM} Alignment Dataset: What Participatory, Representative
               and Individualised Human Feedback Reveals About the Subjective and
               Multicultural Alignment of Large Language Models},
  booktitle = {Advances in Neural Information Processing Systems},
  volume    = {37},
  pages     = {105236--105344},
  year      = {2024},
  publisher = {Neural Information Processing Systems Foundation},
  address   = {Vancouver, BC, Canada},
  doi       = {10.52202/079017-3342},
  url       = {https://proceedings.nips.cc/paper_files/paper/2024/hash/be2e1b68b44f2419e19f6c35a1b8cf35-Abstract-Datasets_and_Benchmarks_Track.html}
}

@inproceedings{Genc_Gu_Degachi_Niforatos_Chandrasegaran_Verma_2026, address={New York, NY, USA}, series={CHI ’26}, title={The Bots of Persuasion: Examining How Conversational Agents’ Linguistic Expressions of Personality Affect User Perceptions and Decisions}, ISBN={979-8-4007-2278-3}, url={https://dl.acm.org/doi/10.1145/3772318.3791407}, DOI={10.1145/3772318.3791407}, abstractNote={Large Language Model-powered conversational agents (CAs) are increasingly capable of projecting sophisticated personalities through language, but how these projections affect users is unclear. We thus examine how CA personalities expressed linguistically affect user decisions and perceptions in the context of charitable giving. In a crowdsourced study, 360 participants interacted with one of eight CAs, each projecting a personality composed of three linguistic aspects: attitude (optimistic/pessimistic), authority (authoritative/submissive), and reasoning (emotional/rational). While the CA’s composite personality did not affect participants’ decisions, it did affect their perceptions and emotional responses. Particularly, participants interacting with pessimistic CAs felt lower emotional state and lower affinity towards the cause, perceived the CA as less trustworthy and less competent, and yet tended to donate more toward the charity. Perceptions of trust, competence, and situational empathy significantly predicted donation decisions. Our findings emphasize the risks CAs pose as instruments of manipulation, subtly influencing user perceptions and decisions.}, booktitle={Proceedings of the 2026 CHI Conference on Human Factors in Computing Systems}, publisher={Association for Computing Machinery}, author={Genç, Hüseyin Uğur and Gu, Heng and Degachi, Chadha and Niforatos, Evangelos and Chandrasegaran, Senthil and Verma, Himanshu}, year={2026}, month=apr, articleno={796}, numpages={25}, collection={CHI ’26} }

@article{Williams-Ceci_Jakesch_Bhat_Kadoma_Zalmanson_Naaman_2026, title={Biased AI writing assistants shift users’ attitudes on societal issues}, volume={12}, DOI={10.1126/sciadv.adw5578}, abstractNote={Artificial intelligence (AI) writing assistants powered by large language models (LLMs) are increasingly used to make autocomplete suggestions to people as they write text. Can these AI writing assistants affect people’s attitudes in this process? In two large-scale preregistered experiments (N = 2582), we exposed participants writing about important societal issues to an AI writing assistant that provided biased autocomplete suggestions. When using the AI assistant, the attitudes participants expressed in a posttask survey converged toward the AI’s position. However, a majority of participants were unaware of the AI suggestions’ bias and their influence. Further, the influence of the AI writing assistant was stronger than the influence of similar suggestions presented as static text, showing that the influence is not fully explained by these suggestions, increasing accessibility of the biased information. Last, warning participants about assistants’ bias before or after exposure does not mitigate the attitude-shift effect.}, number={11}, journal={Science Advances}, publisher={American Association for the Advancement of Science}, author={Williams-Ceci, Sterling and Jakesch, Maurice and Bhat, Advait and Kadoma, Kowe and Zalmanson, Lior and Naaman, Mor}, year={2026}, month=mar, pages={eadw5578} }

@inproceedings{Yeo_Jin_Noh_Shin_Kang_Heo_Chung_Hyun_Han_2026, address={New York, NY, USA}, series={CHI ’26}, title={“Can LLMs Persuade Humans with Deception?”: From a Deceptive Strategy Taxonomy to a Large-Scale Empirical Study}, ISBN={979-8-4007-2278-3}, url={https://dl.acm.org/doi/10.1145/3772318.3791188}, DOI={10.1145/3772318.3791188}, abstractNote={Beyond hallucinations, Large Language Models (LLMs) can craft deceptive arguments that erode users’ critical thinking, posing a significant yet underexamined societal risk. To address this gap, we develop a taxonomy of eight deceptive persuasion strategies by integrating top-down rhetorical theory with a bottom-up analysis of 3,360 AI-generated messages by four LLM families and examining their effects on user perceptions. Through a large-scale user study (N=602) complemented by a think-aloud protocol, we found that participants were vulnerable to Information Manipulation and Uncertainty Exploitation, especially when a message contradicted their prior beliefs. Vulnerability was significantly higher for participants with low cognitive reflection, low topic knowledge, and low topic involvement. Qualitative analyses further revealed that participants were persuaded by the plausibility of an overall narrative even when they distrusted specific details, interpreting deceptive outputs as logically framed information that broadens perspective. We discuss critical implications of these findings for the design of trustworthy AI systems, adaptive user interfaces, and targeted literacy education.}, booktitle={Proceedings of the 2026 CHI Conference on Human Factors in Computing Systems}, publisher={Association for Computing Machinery}, author={Yeo, Haein and Jin, Seungwan and Noh, Taehyung and Shin, Yejin and Kang, Sangyeon and Heo, Sangwoo and Chung, Jiwon and Hyun, Hwarim and Han, Kyungsik}, year={2026}, month=apr, articleno={1506}, numpages={21}, collection={CHI ’26} }

@incollection{schwartz1992universals,
  author    = {Schwartz, Shalom H.},
  title     = {Universals in the Content and Structure of Values: Theoretical
               Advances and Empirical Tests in 20 Countries},
  booktitle = {Advances in Experimental Social Psychology},
  year      = {1992},
  volume    = {25},
  pages     = {1--65},
  publisher = {Elsevier},
  doi       = {10.1016/S0065-2601(08)60281-6}
}

@article{bartneck2009godspeed,
  author  = {Bartneck, Christoph and Kuli{\'c}, Dana and Croft, Elizabeth and
             Zoghbi, Susana},
  title   = {Measurement Instruments for the Anthropomorphism, Animacy,
             Likeability, Perceived Intelligence, and Perceived Safety of Robots},
  journal = {International Journal of Social Robotics},
  year    = {2009},
  volume  = {1},
  number  = {1},
  pages   = {71--81},
  doi     = {10.1007/s12369-008-0001-3}
}

@inproceedings{ullman2019trust,
  author    = {Ullman, Daniel and Malle, Bertram F.},
  title     = {Measuring Gains and Losses in Human-Robot Trust: Evidence for
               Differentiable Components of Trust},
  booktitle = {2019 14th ACM/IEEE International Conference on Human-Robot
               Interaction (HRI)},
  year      = {2019},
  pages     = {618--619},
  publisher = {IEEE},
  address   = {Daegu, South Korea},
  doi       = {10.1109/HRI.2019.8673154}
}

@article{cheng2026sycophantic,
  author  = {Cheng, Myra and Lee, Cinoo and Khadpe, Pranav and Yu, Sunny and
             Han, Dyllan and Jurafsky, Dan},
  title   = {Sycophantic AI Decreases Prosocial Intentions and Promotes
             Dependence},
  journal = {Science},
  year    = {2026},
  volume  = {391},
  number  = {6792},
  pages   = {eaec8352},
  doi     = {10.1126/science.aec8352}
}

@inproceedings{wei2023bot,
  author    = {Wei, Christina Ziying and Kim, Young-Ho and Kuzminykh, Anastasia},
  title     = {The Bot on Speaking Terms: The Effects of Conversation
               Architecture on Perceptions of Conversational Agents},
  booktitle = {Proceedings of the 5th International Conference on Conversational
               User Interfaces},
  series    = {CUI '23},
  year      = {2023},
  articleno = {18},
  numpages  = {16},
  publisher = {Association for Computing Machinery},
  address   = {New York, NY, USA},
  location  = {Eindhoven, Netherlands},
  isbn      = {9798400700149},
  doi       = {10.1145/3571884.3597139}
}

@inproceedings{ortloff2025effectsizes,
  author    = {Ortloff, Anna-Marie and Martius, Florin and Meier, Mischa and
               Raimbault, Theo and Geierhaas, Lisa and Smith, Matthew},
  title     = {Small, Medium, Large? A Meta-Study of Effect Sizes at {CHI} to
               Aid Interpretation of Effect Sizes and Power Calculation},
  booktitle = {Proceedings of the 2025 CHI Conference on Human Factors in
               Computing Systems},
  series    = {CHI '25},
  year      = {2025},
  articleno = {483},
  numpages  = {28},
  publisher = {Association for Computing Machinery},
  address   = {New York, NY, USA},
  doi       = {10.1145/3706598.3713671}
}

@inproceedings{rahman2026vibecheck,
  author = {Rahman, Hasibur and Desai, Smit},
  title = {Vibe Check: Understanding the Effects of LLM-Based Conversational Agents' Personality and Alignment on User Perceptions in Goal-Oriented Tasks},
  year = {2026},
  isbn = {9798400722783},
  publisher = {Association for Computing Machinery},
  address = {New York, NY, USA},
  url = {https://doi.org/10.1145/3772318.3790388},
  doi = {10.1145/3772318.3790388},
  booktitle = {Proceedings of the 2026 CHI Conference on Human Factors in Computing Systems},
  articleno = {371},
  numpages = {30},
  series = {CHI '26}
}

@article{schwartz2012overview,
  author  = {Schwartz, Shalom H.},
  title   = {An Overview of the {Schwartz} Theory of Basic Values},
  journal = {Online Readings in Psychology and Culture},
  year    = {2012},
  volume  = {2},
  number  = {1},
  doi     = {10.9707/2307-0919.1116}
}

@article{schwartz2012refined,
  author  = {Schwartz, Shalom H. and Cieciuch, Jan and Vecchione, Michele
             and Davidov, Eldad and Fischer, Ronald and Beierlein, Constanze
             and Ramos, Alice and Verkasalo, Markku and L{\"o}nnqvist, Jan-Erik
             and Demirutku, Kursad and Dirilen-Gumus, Ozlem and Konty, Mark},
  title   = {Refining the Theory of Basic Individual Values},
  journal = {Journal of Personality and Social Psychology},
  year    = {2012},
  volume  = {103},
  number  = {4},
  pages   = {663--688},
  doi     = {10.1037/a0029393}
}

@article{schwartz2022pvqrr,
  author  = {Schwartz, Shalom H. and Cieciuch, Jan},
  title   = {Measuring the Refined Theory of Individual Values in 49 Cultural
             Groups: Psychometrics of the Revised Portrait Value Questionnaire},
  journal = {Assessment},
  year    = {2022},
  volume  = {29},
  number  = {5},
  pages   = {1005--1019},
  doi     = {10.1177/1073191121998760}
}

@article{russo2022values,
  author  = {Russo, Claudia and Danioni, Francesca and Zagrean, Ioana
             and Barni, Daniela},
  title   = {Changing Personal Values through Value-Manipulation Tasks:
             A Systematic Literature Review Based on {Schwartz}'s Theory of
             Basic Human Values},
  journal = {European Journal of Investigation in Health, Psychology and
             Education},
  year    = {2022},
  volume  = {12},
  number  = {7},
  pages   = {692--715},
  doi     = {10.3390/ejihpe12070052}
}

@article{bardi2011dual,
  author  = {Bardi, Anat and Goodwin, Robin},
  title   = {The Dual Route to Value Change: Individual Processes and Cultural
             Moderators},
  journal = {Journal of Cross-Cultural Psychology},
  year    = {2011},
  volume  = {42},
  number  = {2},
  pages   = {271--287},
  doi     = {10.1177/0022022110396916}
}

@article{arieli2014benevolence,
  author  = {Arieli, Sharon and Grant, Adam M. and Sagiv, Lilach},
  title   = {Convincing Yourself to Care About Others: An Intervention for
             Enhancing Benevolence Values},
  journal = {Journal of Personality},
  year    = {2014},
  volume  = {82},
  number  = {1},
  pages   = {15--24},
  doi     = {10.1111/jopy.12029}
}

@article{ye2019culturalpriming,
  author  = {Ye, Shengquan and Ng, Ting Kin},
  title   = {Value Change in Response to Cultural Priming: The Role of
             Cultural Identity and the Impact on Subjective Well-Being},
  journal = {International Journal of Intercultural Relations},
  year    = {2019},
  volume  = {70},
  pages   = {89--103},
  doi     = {10.1016/j.ijintrel.2019.03.003}
}

@inproceedings{santurkar2023opinions,
  author    = {Santurkar, Shibani and Durmus, Esin and Ladhak, Faisal and
               Lee, Cinoo and Liang, Percy and Hashimoto, Tatsunori},
  title     = {Whose Opinions Do Language Models Reflect?},
  booktitle = {Proceedings of the 40th International Conference on Machine
               Learning},
  year      = {2023},
  volume    = {202},
  series    = {Proceedings of Machine Learning Research},
  pages     = {29971--30004},
  publisher = {PMLR},
  address   = {Honolulu, Hawaii, USA},
  url       = {https://proceedings.mlr.press/v202/santurkar23a.html}
}

@inproceedings{rottger2024politicalcompass,
  author    = {Röttger, Paul and Hofmann, Valentin and Pyatkin, Valentina and
               Hinck, Musashi and Kirk, Hannah and Schuetze, Hinrich and
               Hovy, Dirk},
  title     = {Political Compass or Spinning Arrow? Towards More Meaningful
               Evaluations for Values and Opinions in Large Language Models},
  booktitle = {Proceedings of the 62nd Annual Meeting of the Association for
               Computational Linguistics (Volume 1: Long Papers)},
  year      = {2024},
  pages     = {15295--15311},
  publisher = {Association for Computational Linguistics},
  address   = {Bangkok, Thailand},
  doi       = {10.18653/v1/2024.acl-long.816}
}

@inproceedings{moore2024valueconsistency,
  author    = {Moore, Jared and Deshpande, Tanvi and Yang, Diyi},
  title     = {Are Large Language Models Consistent over Value-Laden
               Questions?},
  booktitle = {Findings of the Association for Computational Linguistics:
               {EMNLP} 2024},
  year      = {2024},
  pages     = {15185--15221},
  publisher = {Association for Computational Linguistics},
  address   = {Miami, Florida, USA},
  doi       = {10.18653/v1/2024.findings-emnlp.891}
}

@inproceedings{yao2024valuefulcra,
  author    = {Yao, Jing and Yi, Xiaoyuan and Gong, Yifan and Wang, Xiting and
               Xie, Xing},
  title     = {Value {FULCRA}: Mapping Large Language Models to the
               Multidimensional Spectrum of Basic Human Value},
  booktitle = {Proceedings of the 2024 Conference of the North American
               Chapter of the Association for Computational Linguistics:
               Human Language Technologies (Volume 1: Long Papers)},
  year      = {2024},
  pages     = {8762--8785},
  publisher = {Association for Computational Linguistics},
  address   = {Mexico City, Mexico},
  doi       = {10.18653/v1/2024.naacl-long.486}
}

@inproceedings{lee2026valueinertia,
  author    = {Lee, Bruce W. and Lee, Yeongheon and Cho, Hyunsoo},
  title     = {Inertia in Moral and Value Judgments of Large Language Models},
  booktitle = {Proceedings of the 64th Annual Meeting of the Association for
               Computational Linguistics (Volume 1: Long Papers)},
  year      = {2026},
  pages     = {27053--27075},
  publisher = {Association for Computational Linguistics},
  address   = {San Diego, California, United States},
  doi       = {10.18653/v1/2026.acl-long.1246}
}

@article{rahman2026ccd,
  author  = {Rahman, Hasibur and Salam, Hanan},
  title   = {{CCD-Bench}: Probing Cultural Conflict in Large Language Model
             Decision-Making},
  journal = {Proceedings of the AAAI Conference on Artificial Intelligence},
  year    = {2026},
  volume  = {40},
  number  = {46},
  pages   = {39125--39133},
  doi     = {10.1609/aaai.v40i46.41260},
  url     = {https://ojs.aaai.org/index.php/AAAI/article/view/41260}
}

@inproceedings{wang2024cdeval,
  author    = {Wang, Yuhang and Zhu, Yanxu and Kong, Chao and Wei, Shuyu and
               Yi, Xiaoyuan and Xie, Xing and Sang, Jitao},
  title     = {{CDE}val: A Benchmark for Measuring the Cultural Dimensions of
               Large Language Models},
  booktitle = {Proceedings of the 2nd Workshop on Cross-Cultural
               Considerations in NLP},
  year      = {2024},
  pages     = {1--16},
  publisher = {Association for Computational Linguistics},
  address   = {Bangkok, Thailand},
  doi       = {10.18653/v1/2024.c3nlp-1.1}
}

@misc{karinshak2024llmglobe,
  author        = {Karinshak, Elise and Hu, Amanda and Kong, Kewen and Rao,
                   Vishwanatha and Wang, Jingren and Wang, Jindong and Zeng, Yi},
  title         = {{LLM-GLOBE}: A Benchmark Evaluating the Cultural Values
                   Embedded in {LLM} Output},
  year          = {2024},
  eprint        = {2411.06032},
  archiveprefix = {arXiv},
  primaryclass  = {cs.CL},
  doi           = {10.48550/arXiv.2411.06032},
  url           = {https://arxiv.org/abs/2411.06032}
}

@inproceedings{alkhamissi2024cultural,
  author    = {AlKhamissi, Badr and ElNokrashy, Muhammad and Alkhamissi, Mai
               and Diab, Mona},
  title     = {Investigating Cultural Alignment of Large Language Models},
  booktitle = {Proceedings of the 62nd Annual Meeting of the Association for
               Computational Linguistics (Volume 1: Long Papers)},
  year      = {2024},
  pages     = {12404--12422},
  publisher = {Association for Computational Linguistics},
  address   = {Bangkok, Thailand},
  doi       = {10.18653/v1/2024.acl-long.671}
}

@article{lu2025cultural,
  author  = {Lu, Jackson G. and Song, Lesley Luyang and Zhang, Lu Doris},
  title   = {Cultural Tendencies in Generative {AI}},
  journal = {Nature Human Behaviour},
  year    = {2025},
  volume  = {9},
  number  = {11},
  pages   = {2360--2369},
  doi     = {10.1038/s41562-025-02242-1}
}

@misc{huang2026valact,
  author = {Huang, Jen-tse and Qin, Jiantong and Qiu, Xueli and Levy, Sharon
            and Kaufman, Michelle R. and Dredze, Mark},
  title  = {Knowing But Not Doing: Convergent Morality and Divergent Action
            in {LLMs}},
  year   = {2026},
  note   = {arXiv:2601.07972},
  url    = {https://arxiv.org/abs/2601.07972}
}

@article{williams1949designs,
  author  = {Williams, E. J.},
  title   = {Experimental Designs Balanced for the Estimation of Residual
             Effects of Treatments},
  journal = {Australian Journal of Scientific Research, Series A},
  year    = {1949},
  volume  = {2},
  number  = {2},
  pages   = {149--168},
  doi     = {10.1071/CH9490149}
}

@article{johnson2023dsi,
  author  = {Johnson, Dan R. and Kaufman, James C. and Baker, Brendan S.
             and Patterson, John D. and Barbot, Baptiste and Green, Adam E.
             and van Hell, Janet and Kennedy, Evan and Sullivan, Grace F.
             and Taylor, Christa L. and Ward, Thomas and Beaty, Roger E.},
  title   = {Divergent Semantic Integration ({DSI}): Extracting Creativity
             from Narratives with Distributional Semantic Modeling},
  journal = {Behavior Research Methods},
  year    = {2023},
  volume  = {55},
  number  = {7},
  pages   = {3726--3759},
  doi     = {10.3758/s13428-022-01986-2}
}

@article{maio1998truisms,
  author  = {Maio, Gregory R. and Olson, James M.},
  title   = {Values as Truisms: Evidence and Implications},
  journal = {Journal of Personality and Social Psychology},
  year    = {1998},
  volume  = {74},
  number  = {2},
  pages   = {294--311},
  doi     = {10.1037/0022-3514.74.2.294}
}

@article{bernard2003introspection,
  author  = {Bernard, Mark M. and Maio, Gregory R. and Olson, James M.},
  title   = {Effects of Introspection About Reasons for Values: Extending
             Research on Values-as-Truisms},
  journal = {Social Cognition},
  year    = {2003},
  volume  = {21},
  number  = {1},
  pages   = {1--25},
  doi     = {10.1521/soco.21.1.1.21193}
}

@inproceedings{reicherts2025extendai,
  author    = {Reicherts, Leon and Zhang, Zelun Tony and von Oswald, Elisabeth
               and Liu, Yuanting and Rogers, Yvonne and Hassib, Mariam},
  title     = {{AI}, Help Me Think---but for Myself: Assisting People in
               Complex Decision-Making by Providing Different Kinds of
               Cognitive Support},
  booktitle = {Proceedings of the 2025 {CHI} Conference on Human Factors
               in Computing Systems},
  series    = {CHI '25},
  year      = {2025},
  publisher = {Association for Computing Machinery},
  address   = {New York, NY, USA},
  articleno = {255},
  numpages = {19},
  doi       = {10.1145/3706598.3713295}
}

@misc{wise2025chatbot,
  author = {Wise, Anthony and Zhou, Xinyi and Reimann, Martin and Dey, Anind
            and Battle, Leilani},
  title  = {A Crowdsourced Study of {ChatBot} Influence in Value-Driven
            Decision Making Scenarios},
  year   = {2025},
  note   = {arXiv:2511.15857},
  url    = {https://arxiv.org/abs/2511.15857}
}

@inproceedings{li2026personality,
  author    = {Li, Jingshu and Song, Tianqi and Boonprakong, Nattapat
               and Zhu, Zicheng and Yang, Yitian and Lee, Yi-Chieh},
  title     = {{AI}-exhibited Personality Traits Can Shape Human Self-concept
               through Conversations},
  booktitle = {Proceedings of the 2026 {CHI} Conference on Human Factors
               in Computing Systems},
  series    = {CHI '26},
  year      = {2026},
  publisher = {Association for Computing Machinery},
  address   = {New York, NY, USA},
  isbn      = {979-8-4007-2278-3},
  pages     = {1--20},
  doi       = {10.1145/3772318.3790654},
  articleno = {8},
  numpages = {20}
}

@article{costello2024conspiracy,
  author  = {Costello, Thomas H. and Pennycook, Gordon and Rand, David G.},
  title   = {Durably Reducing Conspiracy Beliefs through Dialogues with {AI}},
  journal = {Science},
  year    = {2024},
  volume  = {385},
  number  = {6714},
  pages   = {eadq1814},
  doi     = {10.1126/science.adq1814}
}

@inproceedings{sharma2024sycophancy,
  author    = {Sharma, Mrinank and Tong, Meg and Korbak, Tomasz
               and Duvenaud, David and Askell, Amanda and Bowman, Samuel R.
               and Cheng, Newton and Durmus, Esin and Hatfield-Dodds, Zac
               and Johnston, Scott R. and Kravec, Shauna and Maxwell, Timothy
               and McCandlish, Sam and Ndousse, Kamal and Rausch, Oliver
               and Schiefer, Nicholas and Yan, Da and Zhang, Miranda
               and Perez, Ethan},
  title     = {Towards Understanding Sycophancy in Language Models},
  booktitle = {The Twelfth International Conference on Learning
               Representations},
  year      = {2024},
  url       = {https://openreview.net/forum?id=tvhaxkMKAn}
}

@inproceedings{jain2026sycophancy,
  author    = {Jain, Shomik and Park, Charlotte and Viana, Matt
               and Wilson, Ashia and Calacci, Dana},
  title     = {Interaction Context Often Increases Sycophancy in {LLMs}},
  booktitle = {Proceedings of the 2026 {CHI} Conference on Human Factors
               in Computing Systems},
  series    = {CHI '26},
  year      = {2026},
  publisher = {Association for Computing Machinery},
  address   = {New York, NY, USA},
  articleno = {793},
  numpages = {26},
  doi       = {10.1145/3772318.3791915}
}

@inproceedings{jakesch2023cowriting,
  author    = {Jakesch, Maurice and Bhat, Advait and Buschek, Daniel
               and Zalmanson, Lior and Naaman, Mor},
  title     = {Co-Writing with Opinionated Language Models Affects Users'
               Views},
  booktitle = {Proceedings of the 2023 {CHI} Conference on Human Factors
               in Computing Systems},
  series    = {CHI '23},
  year      = {2023},
  articleno = {111},
  numpages  = {15},
  publisher = {Association for Computing Machinery},
  address   = {New York, NY, USA},
  doi       = {10.1145/3544548.3581196}
}

@article{krugel2023moral,
  author  = {Kr{\"u}gel, Sebastian and Ostermaier, Andreas and Uhl, Matthias},
  title   = {{ChatGPT}'s Inconsistent Moral Advice Influences Users'
             Judgment},
  journal = {Scientific Reports},
  year    = {2023},
  volume  = {13},
  number  = {1},
  pages   = {4569},
  doi     = {10.1038/s41598-023-31341-0}
}

@article{glass1976meta,
  author  = {Glass, Gene V.},
  title   = {Primary, Secondary, and Meta-Analysis of Research},
  journal = {Educational Researcher},
  year    = {1976},
  volume  = {5},
  number  = {10},
  pages   = {3--8},
  doi     = {10.3102/0013189X005010003}
}

@article{holm1979multiple,
  author  = {Holm, Sture},
  title   = {A Simple Sequentially Rejective Multiple Test Procedure},
  journal = {Scandinavian Journal of Statistics},
  year    = {1979},
  volume  = {6},
  number  = {2},
  pages   = {65--70}
}

@article{bardi2009structure,
  author  = {Bardi, Anat and Lee, Julie A. and Hofmann-Towfigh, Nadi
             and Soutar, Geoffrey},
  title   = {The Structure of Intraindividual Value Change},
  journal = {Journal of Personality and Social Psychology},
  year    = {2009},
  volume  = {97},
  number  = {5},
  pages   = {913--929},
  doi     = {10.1037/a0016617}
}

@article{lakens2017equivalence,
  author  = {Lakens, Dani{\"e}l},
  title   = {Equivalence Tests: A Practical Primer for t Tests,
             Correlations, and Meta-Analyses},
  journal = {Social Psychological and Personality Science},
  year    = {2017},
  volume  = {8},
  number  = {4},
  pages   = {355--362},
  doi     = {10.1177/1948550617697177}
}

@inproceedings{anderson2024homogenization,
  author    = {Anderson, Barrett R. and Shah, Jash Hemant and
               Kreminski, Max},
  title     = {Homogenization Effects of Large Language Models on Human
               Creative Ideation},
  booktitle = {Proceedings of the 16th Conference on Creativity and
               Cognition},
  series    = {C\&C '24},
  year      = {2024},
  pages     = {413--425},
  publisher = {Association for Computing Machinery},
  address   = {New York, NY, USA},
  doi       = {10.1145/3635636.3656204}
}

@article{danziger2012idealistic,
  author  = {Danziger, Shai and Montal, Ronit and Barkan, Rachel},
  title   = {Idealistic Advice and Pragmatic Choice: A Psychological Distance
             Account},
  journal = {Journal of Personality and Social Psychology},
  year    = {2012},
  volume  = {102},
  number  = {6},
  pages   = {1105--1117},
  doi     = {10.1037/a0027013}
}

@article{verplanken2002motivated,
  author  = {Verplanken, Bas and Holland, Rob W.},
  title   = {Motivated Decision Making: Effects of Activation and
             Self-Centrality of Values on Choices and Behavior},
  journal = {Journal of Personality and Social Psychology},
  year    = {2002},
  volume  = {82},
  number  = {3},
  pages   = {434--447},
  doi     = {10.1037/0022-3514.82.3.434}
}

@article{maio2009changing,
  author  = {Maio, Gregory R. and Pakizeh, Ali and Cheung, Wing-Yee
             and Rees, Kerry J.},
  title   = {Changing, Priming, and Acting on Values: Effects via
             Motivational Relations in a Circular Model},
  journal = {Journal of Personality and Social Psychology},
  year    = {2009},
  volume  = {97},
  number  = {4},
  pages   = {699--715},
  doi     = {10.1037/a0016420}
}

@article{sagiv2017values,
  author  = {Sagiv, Lilach and Roccas, Sonia and Cieciuch, Jan
             and Schwartz, Shalom H.},
  title   = {Personal Values in Human Life},
  journal = {Nature Human Behaviour},
  year    = {2017},
  volume  = {1},
  number  = {9},
  pages   = {630--639},
  doi     = {10.1038/s41562-017-0185-3}
}

@inproceedings{yun2025format,
  author    = {Yun, Longfei and An, Chenyang and Wang, Zilong
               and Peng, Letian and Shang, Jingbo},
  title     = {The Price of Format: Diversity Collapse in {LLMs}},
  booktitle = {Findings of the Association for Computational Linguistics:
               {EMNLP} 2025},
  year      = {2025},
  pages     = {15454--15468},
  address   = {Suzhou, China},
  publisher = {Association for Computational Linguistics},
  doi       = {10.18653/v1/2025.findings-emnlp.836}
}

@article{shrout1979icc,
  author  = {Shrout, Patrick E. and Fleiss, Joseph L.},
  title   = {Intraclass Correlations: Uses in Assessing Rater Reliability},
  journal = {Psychological Bulletin},
  year    = {1979},
  volume  = {86},
  number  = {2},
  pages   = {420--428},
  doi     = {10.1037/0033-2909.86.2.420}
}

@article{covington2010mattr,
  author  = {Covington, Michael A. and McFall, Joe D.},
  title   = {Cutting the {Gordian} Knot: The Moving-Average Type--Token
             Ratio ({MATTR})},
  journal = {Journal of Quantitative Linguistics},
  year    = {2010},
  volume  = {17},
  number  = {2},
  pages   = {94--100},
  doi     = {10.1080/09296171003643098}
}

@inproceedings{seabold2010statsmodels,
  author    = {Seabold, Skipper and Perktold, Josef},
  title     = {Statsmodels: Econometric and Statistical Modeling with {Python}},
  booktitle = {Proceedings of the 9th {Python} in Science Conference},
  year      = {2010},
  pages     = {92--96},
  doi       = {10.25080/Majora-92bf1922-011}
}

@article{hackenburg2025levers,
  author  = {Hackenburg, Kobi and Tappin, Ben M. and Hewitt, Luke and
             Saunders, Ed and Black, Sid and Lin, Hause and Fist, Catherine and
             Margetts, Helen and Rand, David G. and Summerfield, Christopher},
  title   = {The Levers of Political Persuasion with Conversational
             Artificial Intelligence},
  journal = {Science},
  year    = {2025},
  volume  = {390},
  number  = {6777},
  pages   = {eaea3884},
  doi     = {10.1126/science.aea3884}
}

@inproceedings{danry2023askme,
  author    = {Danry, Valdemar and Pataranutaporn, Pat and Mao, Yaoli
               and Maes, Pattie},
  title     = {Don't Just Tell Me, Ask Me: {AI} Systems that Intelligently
               Frame Explanations as Questions Improve Human Logical
               Discernment Accuracy over Causal {AI} Explanations},
  booktitle = {Proceedings of the 2023 {CHI} Conference on Human Factors
               in Computing Systems},
  series    = {CHI '23},
  year      = {2023},
  articleno = {352},
  numpages  = {13},
  publisher = {Association for Computing Machinery},
  address   = {New York, NY, USA},
  doi       = {10.1145/3544548.3580672}
}

@article{mccarthy2010mtld,
  author  = {McCarthy, Philip M. and Jarvis, Scott},
  title   = {{MTLD}, vocd-{D}, and {HD-D}: A Validation Study of Sophisticated
             Approaches to Lexical Diversity Assessment},
  journal = {Behavior Research Methods},
  year    = {2010},
  volume  = {42},
  number  = {2},
  pages   = {381--392},
  doi     = {10.3758/BRM.42.2.381}
}

@article{aitchison1982compositional,
  author  = {Aitchison, John},
  title   = {The Statistical Analysis of Compositional Data},
  journal = {Journal of the Royal Statistical Society, Series B
             (Methodological)},
  year    = {1982},
  volume  = {44},
  number  = {2},
  pages   = {139--160},
  doi     = {10.1111/j.2517-6161.1982.tb01195.x}
}

@article{doshi2024generative,
  author  = {Doshi, Anil R. and Hauser, Oliver P.},
  title   = {Generative {AI} Enhances Individual Creativity but Reduces the
             Collective Diversity of Novel Content},
  journal = {Science Advances},
  year    = {2024},
  volume  = {10},
  number  = {28},
  pages   = {eadn5290},
  doi     = {10.1126/sciadv.adn5290}
}

@article{anderson2006permdisp,
  author  = {Anderson, Marti J.},
  title   = {Distance-Based Tests for Homogeneity of Multivariate Dispersions},
  journal = {Biometrics},
  year    = {2006},
  volume  = {62},
  number  = {1},
  pages   = {245--253},
  doi     = {10.1111/j.1541-0420.2005.00440.x}
}

@article{anderson2013permanova,
  author  = {Anderson, Marti J. and Walsh, Daniel C. I.},
  title   = {{PERMANOVA}, {ANOSIM}, and the {Mantel} Test in the Face of
             Heterogeneous Dispersions: What Null Hypothesis Are You Testing?},
  journal = {Ecological Monographs},
  year    = {2013},
  volume  = {83},
  number  = {4},
  pages   = {557--574},
  doi     = {10.1890/12-2010.1}
}

@article{umbrello2019,
author = {Umbrello, S.},
title = {Lethal Autonomous Weapons: Designing War Machines with Values},
journal = {Delphi - Interdisciplinary Review of Emerging Technologies},
volume = {2},
pages = {30-34},
year  = {2019},
}

@inproceedings{burken2023,
title = {Value Sensitive Design meets Participatory Value Evaluation for autonomous systems in Defence},
author = {Boshuijzen-van Burken, C. and Spruit, S. and Fillerup, L. and Mouter, N.},
year = {2023},
doi = {10.1109/ETHICS57328.2023.10155025},
  booktitle = {2023 IEEE International Symposium on Ethics in Engineering, Science, and Technology (ETHICS)},
  publisher = {IEEE},
  pages = {1--5}
}

@article{vernim2022,
title = {A value sensitive design approach for designing AI-based worker assistance systems in manufacturing},
publisher = {Elsevier},
year = {2022},
author = {Vernim, S. and Bauer, H. and Rauch, E. and Ziegler, Marianne Thejls and Umbrello, S.},
doi = {https://doi.org/10.1016/j.procs.2022.01.248},
  journal = {Procedia Computer Science},
  volume = {200},
  pages = {505--516}
}

@article{sadek2025vsca,
  title        = {The Value-Sensitive Conversational Agent Co-Design Framework},
  author       = {Sadek, M. and Calvo, R. A. and Mougenot, C.},
  journal      = {International Journal of Human–Computer Interaction},
  volume       = {41},
  number       = {15},
  pages        = {9533--9564},
  year         = {2025},
  doi          = {10.1080/10447318.2024.2426737}
}

@incollection{vermaas2020,
author={Vermaas, Pieter E. and Hekkert, Paul and Manders-Huits, No{\"e}mi and Tromp, Nynke},
title="Design Methods in Design for Values",
bookTitle={Handbook of Ethics, Values, and Technological Design: Sources, Theory, Values and Application Domains},
year={2015},
publisher={Springer},
pages={179--201},
doi = {https://doi.org/10.1007/978-94-007-6970-0_10},
  editor = {van den Hoven, Jeroen and Vermaas, Pieter E. and van de Poel, Ibo},
  address = {Dordrecht}
}

@incollection{kroes2020,
author={Kroes, Peter and van de Poel, Ibo},
title="Design for Values and the Definition, Specification, and Operationalization of Values",
bookTitle={Handbook of Ethics, Values, and Technological Design: Sources, Theory, Values and Application Domains},
year={2015},
publisher={Springer},
doi = "https://doi.org/10.1007/978-94-007-6970-0_11",
  pages = {151--178},
  editor = {van den Hoven, Jeroen and Vermaas, Pieter E. and van de Poel, Ibo},
  address = {Dordrecht}
}

@article{poel2020,
  author = {van de Poel, I.},
  title = {Embedding Values in Artificial Intelligence (AI) Systems},
  journal   = {Minds and Machines},
  year = {2020},
  pages = {385--409},
  volume = {30},
  doi = {https://doi.org/10.1007/s11023-020-09537-4}
}

@book{friedmanbook,
author = {Friedman, B. and Hendry, D.},
year = {2019},
title = {Value Sensitive Design: Shaping Technology with Moral Imagination},
publisher = {MIT Press},
doi = {https://doi.org/10.7551/mitpress/7585.001.0001}
}

@incollection{taylor1977,
    author = {Taylor, C.},
    title = {What is human agency?},
    publisher = {Basil Blackwell},
    year = {1977},
    pages = {103--135},
    editor = {Mischel, Theodore},
    booktitle = {The Self: Psychological and Philosophical Issues},
  address = {Oxford}
}

@misc{klingefjord_what_2024,
	title = {What are human values, and how do we align {AI} to them?},
	url = {http://arxiv.org/abs/2404.10636},
	doi = {10.48550/arXiv.2404.10636},
	language = {en},
	urldate = {2026-08-28},
	publisher = {arXiv},
	author = {Klingefjord, Oliver and Lowe, Ryan and Edelman, Joe},
	month = apr,
	year = {2024},
	note = {arXiv:2404.10636 [cs.CY]},
}

@incollection{schwartz2004,
  title        = {Mapping and interpreting cultural differences around the world},
  author       = {Schwartz, S. H.},
  booktitle    = {Comparing cultures, Dimensions of culture in a comparative perspective},
  editor       = {Vinken, H. and Soeters, J. and Ester, P.},
  pages        = {43--73},
  year         = {2004},
  publisher    = {Brill},
  address      = {Leiden, The Netherlands},
  doi = {https://doi.org/10.1163/9789047412977_007}
}

@inproceedings{jakesch2022,
author = {Jakesch, M. and Buçinca, Z. and Amershi, S. and Olteanu, A.},
title = {How Different Groups Prioritize Ethical Values for Responsible AI},
year = {2022},
publisher = {Association for Computing Machinery},
doi = {https://doi.org/10.1145/3531146.3533097},
  booktitle = {Proceedings of the 2022 ACM Conference on Fairness, Accountability, and Transparency},
  pages = {310--323},
  series = {FAccT '22}
}

@article{holstein2019,
author = {Holstein, K. and McLaren, B. and Aleven, V.},
title = {Co-Designing a Real-Time Classroom Orchestration Tool to Support Teacher–AI Complementarity},
year = {2019},
volume = {6},
number = {2},
journal = {Journal of Learning Analytics},
pages = {27–52},
doi = {https://doi.org/10.18608/jla.2019.62.3}
}

@inproceedings{nathan2008,
author = {Nathan, L. and Friedman, B. and Klasnja, P. and Kane, S. and Miller, J.},
title = {Envisioning Systemic Effects on Persons and Society throughout Interactive System Design},
year = {2008},
isbn = {9781605580029},
publisher = {Association for Computing Machinery},
address = {New York, NY, USA},
url = {https://doi.org/10.1145/1394445.1394446},
doi = {10.1145/1394445.1394446},
booktitle = {Proceedings of the 7th ACM Conference on Designing Interactive Systems},
pages = {1–10},
numpages = {10},
location = {Cape Town, South Africa},
series = {DIS '08}
}

@article{gornemann2022,
author = {Görnemann, E. and Spiekermann, S.},
title = {Emotional responses to human values in technology: The case of conversational agents},
journal = {Human–Computer Interaction},
pages = {310--337},
year  = {2024},
doi = {https://doi.org/10.1080/07370024.2022.2136094 },
  volume = {39},
  number = {5--6}
}

@incollection{desmet2015,
author="Desmet, P. and Roeser, S.",
title="Emotions in design for values",
bookTitle={Handbook of Ethics, Values, and Technological Design: Sources, Theory, Values and Application Domains},
year="2015",
publisher="Springer",
pages="203--219",
doi = "https://doi.org/10.1007/978-94-007-6970-0_6",
  editor = {van den Hoven, Jeroen and Vermaas, Pieter E. and van de Poel, Ibo},
  address = {Dordrecht}
}

@inproceedings{sadek2026cui,
author = {Sadek, Malak and Mougenot, C{\'e}line},
title = {Whose Values? Demographic Influences on Perceptions of Conversational AI Value Alignment},
year = {2026},
isbn = {9798400727412},
publisher = {Association for Computing Machinery},
address = {New York, NY, USA},
url = {https://doi.org/10.1145/3816046.3816265},
booktitle = {Proceedings of the 8th ACM Conference on Conversational User Interfaces},
articleno = {83},
numpages = {7},
  doi = {10.1145/3816046.3816265},
  series = {CUI '26}
}

@article{bardi2014,
    author = {Bardi, A. and Buchanan, KE. and Goodwin, R. and Slabu, L. and Robinson, M.},
    title = {Value stability and change during self-chosen life transitions: self-selection versus socialization effects},
    journal = {Journal of Personality and Social Psychology},
    year = {2014},
    volume = {106},
    number = {1},
    pages = {131-147},
    doi = {10.1037/a0034818}
}

@article{sadek2024responsible,
  title        = {Challenges of responsible AI in practice: scoping review and recommended actions},
  author       = {Sadek, Malak and Kallina, Emma and Bohn{\'e}, Thomas and Mougenot, C{\'e}line and Calvo, Rafael A. and Cave, Stephen},
  journal      = {AI \& Society},
  volume       = {40},
  pages        = {199--215},
  year         = {2025},
  doi          = {10.1007/s00146-024-01880-9}
}

@inproceedings{varanasi2023,
    author  = {Varanasi, Rama Adithya and Goyal, Nitesh},
    title   = "``It is currently hodgepodge'': Examining AI/ML Practitioners’ Challenges during Co-production of Responsible AI Values.",
    year = {2023},
    publisher = {Association for Computing Machinery},
    doi = {https://doi.org/10.1145/3544548.3580903},
  booktitle = {Proceedings of the 2023 CHI Conference on Human Factors in Computing Systems},
  articleno = {251},
  numpages = {17},
  series = {CHI '23}
}

@article{palmer2023,
  author    = {Palmer, A. and Schwan, D.},
  title     = {More Process, Less Principles: The Ethics of Deploying AI and Robotics in Medicine},
  journal   = {Cambridge Quarterly of Healthcare Ethics},
  year      = {2024},
  pages = {121--134},
  doi = {https://doi.org/10.1017/S0963180123000087 },
  volume = {33},
  number = {1}
}

@book{wallach2009,
    author = {Wallach, W. and Allen, C.},
    title = {Moral Machines: Teaching Robots Right from Wrong},
    publisher = {Oxford: Oxford University Press},
    year = {2009}
}

@article{vera2019,
  author    = {Liao, Q. Vera and Muller, M.},
  title     = {Enabling Value Sensitive {AI} Systems through Participatory Design Fictions},
   journal   = {Computing Research Repository (CoRR)},
  year      = {2019},
    url = {https://arxiv.org/abs/1912.07381}
}

@inproceedings{janowicz2025,
author = {Janowicz, K. and Liu, Z. and Mai, G. and Wang, Z. and Majic, I. and Fortacz, A. and Mckenzie, G. and Gao, S.},
title = {Whose Truth? Pluralistic Geo-Alignment for (Agentic) AI},
year = {2025},
isbn = {9798400720864},
publisher = {Association for Computing Machinery},
address = {New York, NY, USA},
url = {https://doi.org/10.1145/3748636.3760465},
doi = {10.1145/3748636.3760465},
booktitle = {Proceedings of the 33rd ACM International Conference on Advances in Geographic Information Systems},
pages = {799–803},
numpages = {5},
location = {The Graduate Hotel Minneapolis, Minneapolis, MN, USA},
series = {SIGSPATIAL '25}
}

@inproceedings{shahid2026,
author = {Shahid, Farhana and Zhang, Stella and Vashistha, Aditya},
title = {LLMs Homogenize Values in Constructive Arguments on Value-Laden Topics},
year = {2026},
isbn = {9798400722783},
publisher = {Association for Computing Machinery},
address = {New York, NY, USA},
url = {https://doi.org/10.1145/3772318.3791624},
doi = {10.1145/3772318.3791624},
booktitle = {Proceedings of the 2026 CHI Conference on Human Factors in Computing Systems},
articleno = {397},
numpages = {19},
location = {
},
series = {CHI '26}
}

@inproceedings{collins2024,
title={Modulating Language Model Experiences through Frictions},
author={Collins, K.M. and Chen, V. and Sucholutsky, I. and Kirk, H.R. and Sadek, M. and Sargeant, H. and Talwalkar, A. and Weller, A. and Bhatt, U.},
booktitle={NeurIPS 2024 Workshop on Behavioral Machine Learning},
year={2024},
url={https://openreview.net/forum?id=IlY37cF9ri}
}

@inproceedings{birhane2022,
author = {Birhane, Abeba and Kalluri, Pratyusha and Card, Dallas and Agnew, William and Dotan, Ravit and Bao, Michelle},
title = {The Values Encoded in Machine Learning Research},
year = {2022},
isbn = {9781450393522},
publisher = {Association for Computing Machinery},
address = {New York, NY, USA},
url = {https://doi.org/10.1145/3531146.3533083},
doi = {10.1145/3531146.3533083},
booktitle = {Proceedings of the 2022 ACM Conference on Fairness, Accountability, and Transparency},
pages = {173–184},
numpages = {12},
location = {Seoul, Republic of Korea},
series = {FAccT '22}
}

@article{hagendorff2024,
    author = {Hagendorff, T.},
    title = {Deception abilities emerged in large language models},
    journal = {Proc. Natl. Acad. Sci.},
    volume = {121},
    number = {24},
    year = {2024},
    doi = {https://doi.org/10.1073/pnas.2317967121},
  pages = {e2317967121}
}

@misc{salvi2024,
	title = {On the {Conversational} {Persuasiveness} of {Large} {Language} {Models}: {A} {Randomized} {Controlled} {Trial}},
	copyright = {https://creativecommons.org/licenses/by/4.0/},
	shorttitle = {On the {Conversational} {Persuasiveness} of {Large} {Language} {Models}},
	url = {https://www.researchsquare.com/article/rs-4429707/v1},
	doi = {10.21203/rs.3.rs-4429707/v1},
	language = {en},
	urldate = {2026-08-28},
	publisher = {In Review},
	author = {Salvi, Francesco and Ribeiro, Manoel Horta and Gallotti, Riccardo and West, Robert},
	month = jun,
	year = {2024},
}

@misc{fang2025,
	title = {How {AI} and {Human} {Behaviors} {Shape} {Psychosocial} {Effects} of {Extended} {Chatbot} {Use}: {A} {Longitudinal} {Randomized} {Controlled} {Study}},
	shorttitle = {How {AI} and {Human} {Behaviors} {Shape} {Psychosocial} {Effects} of {Extended} {Chatbot} {Use}},
	url = {http://arxiv.org/abs/2503.17473},
	doi = {10.48550/arXiv.2503.17473},
	language = {en},
	urldate = {2026-08-28},
	publisher = {arXiv},
	author = {Fang, Cathy Mengying and Liu, Auren R. and Danry, Valdemar and Lee, Eunhae and Chan, Samantha W. T. and Pataranutaporn, Pat and Maes, Pattie and Phang, Jason and Lampe, Michael and Ahmad, Lama and Agarwal, Sandhini},
	month = oct,
	year = {2025},
	note = {arXiv:2503.17473 [cs.HC]},
}

@article{tao2024,
  title   = {Cultural bias and cultural alignment of large language models},
  author  = {Tao, Y. and Viberg, O. and Baker, R. S. and Kizilcec, R. F.},
  journal = {PNAS Nexus},
  volume  = {3},
  number  = {9},
  pages   = {pgae346},
  year    = {2024},
  doi     = {10.1093/pnasnexus/pgae346},
  url     = {https://doi.org/10.1093/pnasnexus/pgae346}
}

@misc{kazemi2024,
  title        = {Cultural Fidelity in Large-Language Models: An Evaluation of Online Language Resources as a Driver of Model Performance in Value Representation},
  author       = {Kazemi, S. and Gerhardt, G. and Katz, J. and Kuria, C. I. and Pan, E. and Prabhakar, U.},
  year         = {2024},
  eprint       = {2410.10489},
  archivePrefix= {arXiv},
  primaryClass = {cs.CL},
  url          = {https://arxiv.org/abs/2410.10489}
}

@misc{cao2023,
  title        = {Assessing Cross-Cultural Alignment between ChatGPT and Human Societies: An Empirical Study},
  author       = {Cao, Y. and Zhou, L. and Lee, S. and Cabello, L. and Chen, M. and Hershcovich, D.},
  year         = {2023},
  eprint       = {2303.17466},
  archivePrefix= {arXiv},
  primaryClass = {cs.CL},
  url          = {https://arxiv.org/abs/2303.17466}
}

@inproceedings{sambasivan2021,
  author    = {Sambasivan, Nithya and Kapania, Shivani and Highfill, Hannah and Akrong, Diana and Paritosh, Praveen and Aroyo, Lora M.},
  title     = {''Everyone Wants to Do the Model Work, Not the Data Work'': Data Cascades in High-Stakes AI},
  booktitle = {Proceedings of the 2021 CHI Conference on Human Factors in Computing Systems},
  series    = {CHI '21},
  year      = {2021},
  articleno = {39},
  numpages  = {15},
  publisher = {Association for Computing Machinery},
  address   = {New York, NY, USA},
  doi       = {10.1145/3411764.3445518}
}

@article{hadar2024,
	title = {Assessing the {Alignment} of {Large} {Language} {Models} {With} {Human} {Values} for {Mental} {Health} {Integration}: {Cross}-{Sectional} {Study} {Using} {Schwartz}’s {Theory} of {Basic} {Values}},
	volume = {11},
	issn = {2368-7959},
	shorttitle = {Assessing the {Alignment} of {Large} {Language} {Models} {With} {Human} {Values} for {Mental} {Health} {Integration}},
	url = {https://mental.jmir.org/2024/1/e55988},
	doi = {10.2196/55988},
	language = {en},
	urldate = {2026-08-28},
	journal = {JMIR Mental Health},
	author = {Hadar-Shoval, Dorit and Asraf, Kfir and Mizrachi, Yonathan and Haber, Yuval and Elyoseph, Zohar},
	month = apr,
	year = {2024},
	pages = {e55988},
}

@inproceedings{agarwal2025,
author = {Agarwal, Dhruv and Naaman, Mor and Vashistha, Aditya},
title = {AI Suggestions Homogenize Writing Toward Western Styles and Diminish Cultural Nuances},
year = {2025},
isbn = {9798400713941},
publisher = {Association for Computing Machinery},
address = {New York, NY, USA},
url = {https://doi.org/10.1145/3706598.3713564},
doi = {10.1145/3706598.3713564},
booktitle = {Proceedings of the 2025 CHI Conference on Human Factors in Computing Systems},
articleno = {1117},
numpages = {21},
location = {
},
series = {CHI '25}
}

@article{brennan1996,
	address = {US},
	title = {Conceptual pacts and lexical choice in conversation},
	volume = {22},
	issn = {1939-1285},
	doi = {10.1037/0278-7393.22.6.1482},
	number = {6},
	journal = {Journal of Experimental Psychology: Learning, Memory, and Cognition},
	publisher = {American Psychological Association},
	author = {Brennan, Susan E. and Clark, Herbert H.},
	year = {1996},
	pages = {1482--1493},
}

@article{branigan2010,
	address = {Netherlands},
	title = {Linguistic alignment between people and computers},
	volume = {42},
	issn = {1879-1387},
	doi = {10.1016/j.pragma.2009.12.012},
	number = {9},
	journal = {Journal of Pragmatics},
	publisher = {Elsevier Science},
	author = {Branigan, Holly P. and Pickering, Martin J. and Pearson, Jamie and McLean, Janet F.},
	year = {2010},
	pages = {2355--2368},
}

@misc{naous2024,
  title        = {Having Beer after Prayer? Measuring Cultural Bias in Large Language Models},
  author       = {Naous, T. and Ryan, M. J. and Ritter, A. and Xu, W.},
  year         = {2024},
  eprint       = {2305.14456},
  archivePrefix= {arXiv},
  primaryClass = {cs.CL},
  url          = {https://arxiv.org/abs/2305.14456}
}

@misc{durmus2024,
  title        = {Towards Measuring the Representation of Subjective Global Opinions in Language Models},
  author       = {Durmus, E. and Nguyen, K. and Liao, T. I. and Schiefer, N. and Askell, A. and Bakhtin, A. and Chen, C. and Hatfield-Dodds, Z. and Hernandez, D. and Joseph, N. and Lovitt, L. and McCandlish, S. and Sikder, O. and Tamkin, A. and Thamkul, J. and Kaplan, J. and Clark, J. and Ganguli, D.},
  year         = {2023},
  eprint       = {2306.16388},
  archivePrefix= {arXiv},
  primaryClass = {cs.CL},
  url          = {https://arxiv.org/abs/2306.16388v2}
}

@inproceedings{shelby2023,
author = {Shelby, R. and Rismani, S. and Henne, K. and Moon, AJung and Rostamzadeh, N. and Nicholas, P. and Yilla-Akbari, N. and Gallegos, J. and Smart, A. and Garcia, E. and Virk, G.},
title = {Sociotechnical Harms of Algorithmic Systems: Scoping a Taxonomy for Harm Reduction},
year = {2023},
isbn = {9798400702310},
publisher = {Association for Computing Machinery},
address = {New York, NY, USA},
url = {https://doi.org/10.1145/3600211.3604673},
doi = {10.1145/3600211.3604673},
booktitle = {Proceedings of the 2023 AAAI/ACM Conference on AI, Ethics, and Society},
pages = {723–741},
numpages = {19},
location = {Montr\'{e}al, QC, Canada},
series = {AIES '23}
}

@misc{AIcolonialism,
  title        = {Artificial intelligence is creating a new colonial world order},
  author       = {Hao, Karen},
  year         = {2022},
  howpublished = {MIT Technology Review},
  url = {https://www.technologyreview.com/2022/04/19/1049592/artificial-intelligence-colonialism/}
}

@article{okolo2024,
author = {Okolo, C.T. and Agarwal, D. and Dell, N. and Vashistha, A.},
title = {"If it is easy to understand then it will have value": Examining Perceptions of Explainable AI with Community Health Workers in Rural India},
year = {2024},
issue_date = {April 2024},
publisher = {Association for Computing Machinery},
address = {New York, NY, USA},
volume = {8},
number = {CSCW1},
url = {https://doi.org/10.1145/3637348},
doi = {10.1145/3637348},
journal = {Proc. ACM Hum.-Comput. Interact.},
month = apr,
articleno = {71},
numpages = {28}
}

@article{zhou2025,
  title   = {Should {LLMs} Be {WEIRD}? Exploring {WEIRDness} and Human Rights in Large Language Models},
  author  = {Zhou, K. and Constantinides, M. and Quercia, D.},
  journal = {Proceedings of the AAAI/ACM Conference on AI, Ethics, and Society},
  volume  = {8},
  number  = {3},
  pages   = {2808--2820},
  year    = {2025},
  doi     = {10.1609/aies.v8i3.36761},
  url     = {https://ojs.aaai.org/index.php/AIES/article/view/36761}
}

@inproceedings{irani2010,
author = {Irani, Lilly and Vertesi, Janet and Dourish, Paul and Philip, Kavita and Grinter, Rebecca E.},
title = {Postcolonial computing: a lens on design and development},
year = {2010},
isbn = {9781605589299},
publisher = {Association for Computing Machinery},
address = {New York, NY, USA},
url = {https://doi.org/10.1145/1753326.1753522},
doi = {10.1145/1753326.1753522},
booktitle = {Proceedings of the SIGCHI Conference on Human Factors in Computing Systems},
pages = {1311–1320},
numpages = {10},
location = {Atlanta, Georgia, USA},
series = {CHI '10}
}

@inproceedings{safir025,
	title = {Subaltern {Futures} in {AI}},
	url = {https://www.scienceopen.com/hosted-document?doi=10.14236/ewic/BCSHCI2025.29},
	doi = {10.14236/ewic/BCSHCI2025.29},
	language = {en},
	urldate = {2026-08-28},
	publisher = {BCS Learning \& Development},
	author = {Safir, Abdullah Hasan and Hollanek, Tomasz and Blackwell, Alan F. and Debnath, Ramit},
	month = nov,
	year = {2025},
	pages = {298--303},
  booktitle = {BCS HCI 2025: Human Centred Approaches and their Impact on AI System Design, Application, and Evaluation}
}

@misc{oyemike2025,
	title = {Alignment {Debt}: {The} {Hidden} {Work} of {Making} {AI} {Usable}},
	shorttitle = {Alignment {Debt}},
	url = {http://arxiv.org/abs/2511.09663},
	doi = {10.48550/arXiv.2511.09663},
	language = {en},
	urldate = {2026-08-28},
	publisher = {arXiv},
	author = {Oyemike, Cumi and Akpan, Elizabeth and Hervé-Berdys, Pierre},
	month = nov,
	year = {2025},
	note = {arXiv:2511.09663 [cs.CY]},
}

@ARTICLE{nass2000,
  author={Nass, C. and Moon, Y.},
  journal={Journal of Social Issues}, 
  title={Machines and Mindlessness: Social Responses to Computers},
  year={2000},
  volume = {56},
  number = {1},
  pages = {81--103},
  doi = {https://doi.org/10.1111/0022-4537.00153}
}

@book{reeves1996,
	address = {New York, NY, US},
	series = {The media equation:  {How} people treat computers, television, and new media like real people and places},
	title = {The media equation:  {How} people treat computers, television, and new media like real people and places},
	isbn = {978-1-57586-052-7},
	shorttitle = {The media equation},
	publisher = {Cambridge University Press},
	author = {Reeves, Byron and Nass, Clifford Ivar},
	year = {1996},
	note = {Pages: xiv, 305},
}

@article{khadpe2020,
	title = {Conceptual {Metaphors} {Impact} {Perceptions} of {Human}-{AI} {Collaboration}},
	volume = {4},
	issn = {2573-0142},
	url = {http://arxiv.org/abs/2008.02311},
	doi = {10.1145/3415234},
	language = {en},
	number = {CSCW2},
	urldate = {2026-08-28},
	journal = {Proceedings of the ACM on Human-Computer Interaction},
	author = {Khadpe, Pranav and Krishna, Ranjay and Fei-Fei, Li and Hancock, Jeffrey and Bernstein, Michael},
	month = oct,
	year = {2020},
	note = {arXiv:2008.02311 [cs.HC]},
	pages = {1--26},
  articleno = {163},
  numpages = {26}
}

@article{desai2023,
author = {Desai, Smit and Twidale, Michael},
title = {Metaphors in Voice User Interfaces: A Slippery Fish},
year = {2023},
issue_date = {December 2023},
publisher = {Association for Computing Machinery},
address = {New York, NY, USA},
volume = {30},
number = {6},
issn = {1073-0516},
url = {https://doi.org/10.1145/3609326},
doi = {10.1145/3609326},
journal = {ACM Trans. Comput.-Hum. Interact.},
month = sep,
articleno = {89},
numpages = {37}
}

@inproceedings{desai2024,
author = {Desai, Smit and Dubiel, Mateusz and Leiva, Luis A.},
title = {Examining Humanness as a Metaphor to Design Voice User Interfaces},
year = {2024},
isbn = {9798400705113},
publisher = {Association for Computing Machinery},
address = {New York, NY, USA},
url = {https://doi.org/10.1145/3640794.3665535},
doi = {10.1145/3640794.3665535},
booktitle = {Proceedings of the 6th ACM Conference on Conversational User Interfaces},
articleno = {7},
numpages = {15},
location = {Luxembourg, Luxembourg},
series = {CUI '24}
}

@inproceedings{desai2026,
author = {Desai, Smit and Chin, Jessie and Wang, Dakuo and Cowan, Benjamin R. and Twidale, Michael},
title = {Toward Metaphor-Fluid Conversation Design for Voice User Interfaces},
year = {2026},
isbn = {9798400727412},
publisher = {Association for Computing Machinery},
address = {New York, NY, USA},
url = {https://doi.org/10.1145/3816046.3816223},
booktitle = {Proceedings of the 8th ACM Conference on Conversational User Interfaces},
articleno = {19},
numpages = {21},
  doi = {10.1145/3816046.3816223},
  series = {CUI '26}
}

@inproceedings{li2025confidence,
author = {Li, Jingshu and Yang, Yitian and Liao, Q. Vera and Zhang, Junti and Lee, Yi-Chieh},
title = {As Confidence Aligns: Understanding the Effect of AI Confidence on Human Self-confidence in Human-AI Decision Making},
year = {2025},
isbn = {9798400713941},
publisher = {Association for Computing Machinery},
address = {New York, NY, USA},
url = {https://doi.org/10.1145/3706598.3713336},
doi = {10.1145/3706598.3713336},
booktitle = {Proceedings of the 2025 CHI Conference on Human Factors in Computing Systems},
articleno = {1111},
numpages = {16},
location = {
},
series = {CHI '25}
}

@article{chong2007,
	title = {Framing {Theory}},
	volume = {10},
	issn = {1094-2939, 1545-1577},
	url = {https://www.annualreviews.org/doi/10.1146/annurev.polisci.10.072805.103054},
	doi = {10.1146/annurev.polisci.10.072805.103054},
	language = {en},
	number = {1},
	urldate = {2026-08-28},
	journal = {Annual Review of Political Science},
	author = {Chong, Dennis and Druckman, James N.},
	month = jun,
	year = {2007},
	pages = {103--126},
}

@article{lourie2021,
	title = {{SCRUPLES}: {A} {Corpus} of {Community} {Ethical} {Judgments} on 32,000 {Real}-{Life} {Anecdotes}},
	volume = {35},
	copyright = {Copyright (c) 2021 Association for the Advancement of Artificial Intelligence},
	issn = {2374-3468},
	shorttitle = {{SCRUPLES}},
	url = {https://ojs.aaai.org/index.php/AAAI/article/view/17589},
	doi = {10.1609/aaai.v35i15.17589},
	language = {en},
	number = {15},
	urldate = {2026-08-28},
	journal = {Proceedings of the AAAI Conference on Artificial Intelligence},
	author = {Lourie, Nicholas and Bras, Ronan Le and Choi, Yejin},
	month = may,
	year = {2021},
	pages = {13470--13479},
}

@inproceedings{forbes2020,
	address = {Online},
	title = {Social {Chemistry} 101: {Learning} to {Reason} about {Social} and {Moral} {Norms}},
	shorttitle = {Social {Chemistry} 101},
	url = {https://aclanthology.org/2020.emnlp-main.48/},
	doi = {10.18653/v1/2020.emnlp-main.48},
	urldate = {2026-08-28},
	booktitle = {Proceedings of the 2020 {Conference} on {Empirical} {Methods} in {Natural} {Language} {Processing} ({EMNLP})},
	publisher = {Association for Computational Linguistics},
	author = {Forbes, Maxwell and Hwang, Jena D. and Shwartz, Vered and Sap, Maarten and Choi, Yejin},
	editor = {Webber, Bonnie and Cohn, Trevor and He, Yulan and Liu, Yang},
	month = nov,
	year = {2020},
	pages = {653--670},
}

@article{trope2010,
	address = {US},
	title = {Construal-level theory of psychological distance},
	volume = {117},
	issn = {1939-1471},
	doi = {10.1037/a0018963},
	number = {2},
	journal = {Psychological Review},
	publisher = {American Psychological Association},
	author = {Trope, Yaacov and Liberman, Nira},
	year = {2010},
	pages = {440--463},
}

@article{leemediator2020,
author = {Lee, Yi-Chieh and Yamashita, Naomi and Huang, Yun},
title = {Designing a Chatbot as a Mediator for Promoting Deep Self-Disclosure to a Real Mental Health Professional},
year = {2020},
issue_date = {May 2020},
publisher = {Association for Computing Machinery},
address = {New York, NY, USA},
volume = {4},
number = {CSCW1},
url = {https://doi.org/10.1145/3392836},
doi = {10.1145/3392836},
journal = {Proc. ACM Hum.-Comput. Interact.},
month = may,
articleno = {31},
numpages = {27}
}

@inproceedings{leedisclosure2020,
author = {Lee, Yi-Chieh and Yamashita, Naomi and Huang, Yun and Fu, Wai},
title = {"I Hear You, I Feel You": Encouraging Deep Self-disclosure through a Chatbot},
year = {2020},
isbn = {9781450367080},
publisher = {Association for Computing Machinery},
address = {New York, NY, USA},
url = {https://doi.org/10.1145/3313831.3376175},
doi = {10.1145/3313831.3376175},
booktitle = {Proceedings of the 2020 CHI Conference on Human Factors in Computing Systems},
pages = {1–12},
numpages = {12},
location = {Honolulu, HI, USA},
series = {CHI '20}
}

@article{riedl2026,
author = {Riedl, Christoph and Savage, Saiph and Zvelebilova, Josie},
title = {Cognitive Spillover in Human–AI Teams},
year = {2026},
issue_date = {June 2026},
publisher = {Association for Computing Machinery},
address = {New York, NY, USA},
volume = {33},
number = {3},
issn = {1073-0516},
url = {https://doi.org/10.1145/3805039},
doi = {10.1145/3805039},
journal = {ACM Trans. Comput.-Hum. Interact.},
month = jun,
articleno = {32},
numpages = {33}
}

@article{harrell2025,
	title = {Evidence of spillovers from (non)cooperative human-bot to human-human interactions},
	volume = {28},
	issn = {2589-0042},
	url = {https://www.cell.com/iscience/abstract/S2589-0042(25)01267-2},
	doi = {10.1016/j.isci.2025.113006},
	language = {English},
	number = {8},
	urldate = {2026-08-28},
	journal = {iScience},
	publisher = {Elsevier},
	author = {Harrell, Ashley and Traeger, Margaret L.},
	month = aug,
	year = {2025},
  pages = {113006}
}

@article{proteus2007,
	title = {The {Proteus} {Effect}: {The} {Effect} of {Transformed} {Self}-{Representation} on {Behavior}},
	volume = {33},
	issn = {0360-3989},
	shorttitle = {The {Proteus} {Effect}},
	url = {https://doi.org/10.1111/j.1468-2958.2007.00299.x},
	doi = {10.1111/j.1468-2958.2007.00299.x},
	number = {3},
	urldate = {2026-08-28},
	journal = {Human Communication Research},
	author = {Yee, Nick and Bailenson, Jeremy},
	month = jul,
	year = {2007},
	pages = {271--290},
}

@book{kekes1993,
	title = {The {Morality} of {Pluralism}},
	url = {https://www.jstor.org/stable/j.ctt7smh7},
	doi = {10.2307/j.ctt7smh7},
	urldate = {2026-08-28},
	publisher = {Princeton University Press},
	author = {Kekes, John},
	year = {1993},
}

@techreport{chatterji2025,
	title = {How {People} {Use} {ChatGPT}},
	language = {en},
	author = {Chatterji, Aaron and Cunningham, Thomas and Deming, David J and Hitzig, Zoe and Ong, Christopher and Shan, Carl Yan and Wadman, Kevin},
    year = {2025},
  institution = {National Bureau of Economic Research},
  type = {Working Paper},
  number = {34255},
  doi = {10.3386/w34255},
  url = {https://www.nber.org/papers/w34255}
}

@article{henrich2010,
	title = {The weirdest people in the world?},
	volume = {33},
	issn = {1469-1825},
	doi = {10.1017/S0140525X0999152X},
	language = {eng},
	number = {2-3},
	journal = {The Behavioral and Brain Sciences},
	author = {Henrich, Joseph and Heine, Steven J. and Norenzayan, Ara},
	month = jun,
	year = {2010},
	pages = {61--83},
}

@article{veselovsky2025,
author = {Veselovsky, Veniamin and Horta Ribeiro, Manoel and Cozzolino, Philip J. and Gordon, Andrew and Rothschild, David and West, Robert},
title = {
Prevalence and Prevention of Large Language Model Use in Crowd Work},
year = {2025},
issue_date = {March 2025},
publisher = {Association for Computing Machinery},
address = {New York, NY, USA},
volume = {68},
number = {3},
issn = {0001-0782},
url = {https://doi.org/10.1145/3685527},
doi = {10.1145/3685527},
journal = {Commun. ACM},
month = feb,
pages = {42–47},
numpages = {6}
}

@article{kleinberg2021,
    author = {Kleinberg, J. and Raghavan, M.},
    title = {Algorithmic monoculture and social welfare},
    journal = {Proceedings of the National Academy of Sciences},
    year = {2021},
    volume = {118},
    number = {22},
    doi = {https://doi.org/10.1073/pnas.2018340118},
  pages = {e2018340118}
}

@article{gallegos2026,
	title = {Labeling messages as {AI}-generated does not reduce their persuasive effects},
	volume = {5},
	issn = {2752-6542},
	url = {https://doi.org/10.1093/pnasnexus/pgag008},
	doi = {10.1093/pnasnexus/pgag008},
	number = {2},
	urldate = {2026-08-28},
	journal = {PNAS Nexus},
	author = {Gallegos, Isabel O and Shani, Chen and Shi, Weiyan and Bianchi, Federico and Gainsburg, Izzy and Jurafsky, Dan and Willer, Robb},
	month = feb,
	year = {2026},
	pages = {pgag008},
}

@article{higgins1977category,
  author  = {Higgins, E. Tory and Rholes, William S. and Jones, Carl R.},
  title   = {Category Accessibility and Impression Formation},
  journal = {Journal of Experimental Social Psychology},
  year    = {1977},
  volume  = {13},
  number  = {2},
  pages   = {141--154},
  doi     = {10.1016/S0022-1031(77)80007-3}
}

@article{srull1979accessibility,
  author  = {Srull, Thomas K. and Wyer, Robert S.},
  title   = {The Role of Category Accessibility in the Interpretation of
             Information about Persons: Some Determinants and Implications},
  journal = {Journal of Personality and Social Psychology},
  year    = {1979},
  volume  = {37},
  number  = {10},
  pages   = {1660--1672},
  doi     = {10.1037/0022-3514.37.10.1660}
}

@article{higgins1985priming,
  author  = {Higgins, E. Tory and Bargh, John A. and Lombardi, Wendy J.},
  title   = {Nature of Priming Effects on Categorization},
  journal = {Journal of Experimental Psychology: Learning, Memory, and
             Cognition},
  year    = {1985},
  volume  = {11},
  number  = {1},
  pages   = {59--69},
  doi     = {10.1037/0278-7393.11.1.59}
}

@article{hong2000multicultural,
  author  = {Hong, Ying-yi and Morris, Michael W. and Chiu, Chi-yue and
             Benet-Mart{\'i}nez, Ver{\'o}nica},
  title   = {Multicultural Minds: A Dynamic Constructivist Approach to
             Culture and Cognition},
  journal = {American Psychologist},
  year    = {2000},
  volume  = {55},
  number  = {7},
  pages   = {709--720},
  doi     = {10.1037/0003-066X.55.7.709}
}

@article{higgins1978saying,
  author  = {Higgins, E. Tory and Rholes, William S.},
  title   = {``Saying Is Believing'': Effects of Message Modification on
             Memory and Liking for the Person Described},
  journal = {Journal of Experimental Social Psychology},
  year    = {1978},
  volume  = {14},
  number  = {4},
  pages   = {363--378},
  doi     = {10.1016/0022-1031(78)90032-X}
}

@incollection{bem1972selfperception,
  author    = {Bem, Daryl J.},
  title     = {Self-Perception Theory},
  booktitle = {Advances in Experimental Social Psychology},
  editor    = {Berkowitz, Leonard},
  volume    = {6},
  year      = {1972},
  pages     = {1--62},
  publisher = {Academic Press},
  address   = {New York, NY, USA},
  doi       = {10.1016/S0065-2601(08)60024-6}
}

@article{pickering2004dialogue,
  author  = {Pickering, Martin J. and Garrod, Simon},
  title   = {Toward a Mechanistic Psychology of Dialogue},
  journal = {Behavioral and Brain Sciences},
  year    = {2004},
  volume  = {27},
  number  = {2},
  pages   = {169--190},
  doi     = {10.1017/S0140525X04000056}
}

@inproceedings{huang2025valueswild,
  author    = {Huang, Saffron and Durmus, Esin and McCain, Miles and
               Handa, Kunal and Tamkin, Alex and Hong, Jerry and
               Stern, Michael and Somani, Arushi and Zhang, Xiuruo and
               Ganguli, Deep},
  title     = {Values in the Wild: Discovering and Analyzing Values in
               Real-World Language Model Interactions},
  booktitle = {Proceedings of the Second Conference on Language Modeling},
  year      = {2025},
  month     = oct,
  url       = {https://openreview.net/forum?id=zJHZJClG1Z}
}

@misc{teng2026moral,
  author        = {Teng, Yue and Zhong, Qianer and
                   Thordsen, Kim Mai Tich Nguyen and Montag, Christian and
                   Becker, Benjamin},
  title         = {Brief Chatbot Interactions Produce Lasting Changes in
                   Human Moral Values},
  year          = {2026},
  month         = apr,
  eprint        = {2604.21430},
  archivePrefix = {arXiv},
  primaryClass  = {cs.AI},
  doi           = {10.48550/arXiv.2604.21430},
  url           = {https://arxiv.org/abs/2604.21430},
  note          = {arXiv:2604.21430v1}
}

@inproceedings{reicherts2022proberbot,
  author = {Reicherts, Leon and Park, Gun Woo and Rogers, Yvonne},
  title = {Extending Chatbots to Probe Users: Enhancing Complex Decision-Making Through Probing Conversations},
  booktitle = {Proceedings of the 4th Conference on Conversational User Interfaces},
  series = {CUI '22},
  year = {2022},
  articleno = {2},
  numpages = {10},
  publisher = {Association for Computing Machinery},
  doi = {10.1145/3543829.3543832},
  url = {https://dl.acm.org/doi/10.1145/3543829.3543832}
}

@article{khadar2025socratic,
  author = {Khadar, Malik and Runningen, Daniel and Tang, Julia and Chancellor, Stevie and Kaur, Harmanpreet},
  title = {Wisdom of the Crowd, Without the Crowd: A Socratic {LLM} for Asynchronous Deliberation on Perspectivist Data},
  journal = {Proceedings of the ACM on Human-Computer Interaction},
  volume = {9},
  number = {7},
  articleno = {CSCW526},
  numpages = {35},
  year = {2025},
  doi = {10.1145/3757707},
  url = {https://dl.acm.org/doi/10.1145/3757707}
}

@inproceedings{tarvirdians2026reflectimate,
  author = {Tarvirdians, Morita and Chandrasegaran, Senthil and Hung, Hayley and Jonker, Catholijn M. and Oertel, Catharine},
  title = {Reflecti-Mate: A Conversational Agent for Adaptive Decision-Making Support Through System 1 and System 2 Thinking},
  booktitle = {Proceedings of the 34th ACM Conference on User Modeling, Adaptation and Personalization},
  series = {UMAP '26},
  pages = {213--222},
  year = {2026},
  publisher = {Association for Computing Machinery},
  doi = {10.1145/3774935.3806176},
  url = {https://dl.acm.org/doi/10.1145/3774935.3806176}
}

@article{collins2024thoughtpartners,
  author = {Collins, Katherine M. and Sucholutsky, Ilia and Bhatt, Umang and Chandra, Kartik and Wong, Lionel and Lee, Mina and Zhang, Cedegao E. and Zhi-Xuan, Tan and Ho, Mark and Mansinghka, Vikash and Weller, Adrian and Tenenbaum, Joshua B. and Griffiths, Thomas L.},
  title = {Building Machines that Learn and Think with People},
  journal = {Nature Human Behaviour},
  volume = {8},
  pages = {1851--1863},
  year = {2024},
  doi = {10.1038/s41562-024-01991-9},
  url = {https://doi.org/10.1038/s41562-024-01991-9},
  number = {10}
}

@article{morrin2026delusions,
  author = {Morrin, H. and Nicholls, L. and Levin, M. and Yiend, J. and
            Iyengar, U. and DelGuidice, F. and Bhattacharya, S. and Tognin, S. and
            MacCabe, J. and Twumasi, R. and Alderson-Day, B. and Pollak, T. A.},
  title = {Delusions by design? How everyday {AIs} might be fuelling psychosis
           (and what can be done about it)},
  journal = {European Psychiatry},
  volume = {69},
  number = {S1},
  pages = {S113},
  year = {2026},
  note = {Conference abstract, 34th European Congress of Psychiatry},
  doi = {10.1192/j.eurpsy.2026.10633},
  url = {https://pmc.ncbi.nlm.nih.gov/articles/PMC13443459/}
}


\appendix

\section{Supporting Analyses}
\label{app:supporting}

\begin{table}[h]
\centering
\caption{Parallel-form validation at the four higher-order dimensions, from the
pre-study ($N = 100$, full 57-item administration, no manipulation). Cronbach's
$\alpha$ is for the full-instrument score. The correlation column gives the range
across the three blocks of each block's correlation with the corresponding
full-instrument score. ICC(2,1) treats one block as an interchangeable measure and
ICC(2,3) the three together. Block gap is the largest mean difference between any
two blocks, in raw scale points. Equivalence counts the tests passed at a $\pm 0.5$
margin: three block-to-block pairs and three block-to-full comparisons per
dimension, 24 in total.}
\label{tab:parallel-forms}
\small
\begin{tabular}{@{}lrrrrrr@{}}
\toprule
Higher-order dimension & $\alpha$ & $r$ with full & ICC(2,1) & ICC(2,3) & Block gap & Equiv. \\
\midrule
Openness to Change & .86 & .88--.92 & .66 & .85 & 0.37 & 6/6 \\
Self-Enhancement   & .88 & .91--.92 & .73 & .89 & 0.37 & 6/6 \\
Conservation       & .91 & .95--.97 & .84 & .94 & 0.26 & 6/6 \\
Self-Transcendence & .93 & .93--.96 & .83 & .94 & 0.05 & 6/6 \\
\bottomrule
\end{tabular}
\end{table}

\begin{table}[h]
\centering
\caption{Robustness of the Phase-3 spread contraction. Panel A permutes the 200
condition labels and rebuilds every leave-one-out centroid under the shuffled labels, so
the dependence between scores sharing a centroid is reproduced in the null rather than
assumed away; 5{,}000 permutations per condition. Panel B replaces the spread estimator
with five alternatives, each a different definition of spread rather than a variation on
one. Panel C varies what the contrast conditions on.}
\label{tab:rq2-robustness}
\small
\begin{tabular}{@{}lrrr@{}}
\toprule
\multicolumn{4}{@{}l}{\textbf{A. Randomization inference}} \\
Contrast & Estimate & Perm.\ $p$ & Holm (six) \\
\midrule
P2 \arm{ChatGPT} $-$ \arm{Baseline} & $-0.070$ & .598 & .828 \\
P2 \arm{Claude} $-$ \arm{Baseline}  & $-0.109$ & .414 & .828 \\
P2 \arm{Gemini} $-$ \arm{Baseline}  & $-0.274$ & .040 & .159 \\
P3 \arm{ChatGPT} $-$ \arm{Baseline} & $\mathbf{-0.352}$ & $\mathbf{.0036}$ & $\mathbf{.018}$ \\
P3 \arm{Claude} $-$ \arm{Baseline}  & $-0.220$ & .067 & .202 \\
P3 \arm{Gemini} $-$ \arm{Baseline}  & $\mathbf{-0.432}$ & $\mathbf{.0002}$ & $\mathbf{.0012}$ \\
\midrule
\multicolumn{4}{@{}l}{\textbf{B. Alternative definitions of spread}} \\
Estimator & P2 est.\ ($p$) & \multicolumn{2}{r}{P3 est.\ ($p$)} \\
\midrule
Distance to centroid, $L_2$ & $-0.151$ (.155) & \multicolumn{2}{r}{$-0.334$ (\textbf{.0005})} \\
Distance to centroid, $L_1$ & $-0.308$ (.107) & \multicolumn{2}{r}{$-0.611$ (\textbf{.0005})} \\
Distance to spatial median  & $-0.165$ (.123) & \multicolumn{2}{r}{$-0.338$ (\textbf{.0010})} \\
Median, not mean, over people & $-0.205$ (.057) & \multicolumn{2}{r}{$-0.195$ (\textbf{.039})} \\
Mean pairwise distance      & $-0.236$ (.119) & \multicolumn{2}{r}{$-0.455$ (\textbf{.0005})} \\
Log ratio of spread, scale-free & $-0.140$ (.111) & \multicolumn{2}{r}{$-0.267$ (\textbf{.0005})} \\
\midrule
\multicolumn{4}{@{}l}{\textbf{C. What the contrast conditions on}} \\
Check & \multicolumn{3}{r}{P3 result} \\
\midrule
Conditioning on Phase-1 spread & \multicolumn{3}{r}{$-0.268$, $p = .0009$} \\
Uncentred higher-order scores & \multicolumn{3}{r}{$-0.290$, $p = .0090$} \\
Scale use alone (MRAT) & \multicolumn{3}{r}{$-0.003$, $p = .965$} \\
Adjusted for each person's MRAT change & \multicolumn{3}{r}{$-0.296$, $p = .0015$} \\
Dropping any single participant & \multicolumn{3}{r}{$[-0.362, -0.299]$} \\
\bottomrule
\end{tabular}
\end{table}

\begin{table}[h]
\centering
\caption{Perception of the conversational partner, by condition. The three conversational conditions are shown because RQ3 concerns them. Cronbach's $\alpha$ is reported for the multi-item measures. No measure separates the three conditions after Holm correction across the seven measures.}
\label{tab:perception}
\small
\begin{tabular}{@{}llrrrr@{}}
\toprule
Measure & Scale & $\alpha$ & \arm{ChatGPT} & \arm{Claude} & \arm{Gemini} \\
\midrule
Moral trust: ethical      & 0--7 & .95 & 4.99 & 4.95 & 4.83 \\
Moral trust: sincere      & 0--7 & .94 & 4.82 & 4.58 & 4.62 \\
Moral trust (all 8)       & 0--7 & .97 & 4.90 & 4.76 & 4.72 \\
Perf.\ trust: reliable    & 0--7 & .78 & 5.01 & 4.96 & 4.80 \\
Perf.\ trust: capable     & 0--7 & .94 & 5.04 & 5.11 & 4.86 \\
Perf.\ trust (all 8)      & 0--7 & .93 & 5.03 & 5.04 & 4.83 \\
Likeability               & 1--5 & .92 & 3.89 & 4.02 & 3.64 \\
Intelligence              & 1--5 & .92 & 4.13 & 4.11 & 3.98 \\
Response quality          & 1--7 & --- & 5.40 & 5.82 & 5.20 \\
Rightness judgment        & 1--7 & --- & 5.34 & 5.60 & 4.90 \\
Return likelihood         & 1--7 & --- & 4.60 & 4.74 & 4.34 \\
\bottomrule
\end{tabular}
\end{table}

\begin{table}[h]
\centering
\caption{Balance, estimator agreement, and integrity checks. The analysis of covariance
row matters because change scores and covariate adjustment disagree whenever starting
points are unbalanced, so their agreement here is informative rather than redundant.}
\label{tab:checks}
\small
\begin{tabular}{@{}ll@{}}
\toprule
Check & Result \\
\midrule
Phase-1 axis balanced across conditions & $F(3, 196) = 0.45$, $p = .721$ \\
Phase-1 spread equal across conditions & Levene $p = .622$ \\
Participant-typed turns comparable & $F(2, 147) = 0.62$, $p = .541$ \\
Advice length, words typed, words per turn & smallest Holm $p = .585$ \\
ANCOVA agrees with change scores at P2 & $+0.452$ (.0005), $+0.378$ (.0032), $+0.392$ (.0023) \\
\quad\textit{(\arm{ChatGPT} / \arm{Claude} / \arm{Gemini})} & \\
Model words, first turn (identical prompt) & $F(2, 147) = 267.47$, $p < .001$ \\
Model turns & $F(2, 147) = 0.62$, $p = .541$ \\
Words received, as covariate on the P2 shift & $-0.030$ per 100 words, SE $0.035$, $p = .387$ (MDE $0.098$) \\
Provider omnibus, before / after that covariate & $F(2, 144) = 0.18$, $p = .838$ / $F(2, 143) = 0.17$, $p = .846$ \\
Within-condition model uniformity & $0.895$ / $0.886$ / $0.882$ \\
\quad\textit{(\arm{ChatGPT} / \arm{Claude} / \arm{Gemini})} & \\
Centring identity residual & $\le 1.2 \times 10^{-14}$ \\
\bottomrule
\end{tabular}
\end{table}

\section{Relationship to the Preregistration}
\label{app:prereg}

The four research questions stated in \S\ref{sec:intro} are reworded from the
preregistered versions for presentation. They map one-to-one onto the registered questions, in the same order, and they cover the same four conditions, the same three phases, and the same outcomes.

The rewording is presentational. It names the non-interactive AI control in RQ1, states RQ2 as the contrast between a change in spread and a change in direction, and folds the registered sub-questions of RQ3 and RQ4 into single sentences rather than listing each component. No registered question was dropped, and none was added.


\end{document}